\documentclass[,%
reprint,
 amsmath,amssymb,
 aps,
 twocolumn
]{revtex4}

\usepackage [autostyle, english = american]{csquotes}
\MakeOuterQuote{"}

\usepackage[colorlinks=true]{hyperref}

\hypersetup{colorlinks,citecolor=blue,linkcolor=blue,urlcolor=blue}
\hypersetup{final=true}
\usepackage{aas_macros}
\usepackage{todonotes}
\usepackage{graphicx}
\usepackage{dcolumn}
\usepackage{bm}

\begin{document}


\title{Inhomogeneous Cosmology with Numerical Relativity}

\author{Hayley J. Macpherson}
 \email{hayley.macpherson@monash.edu}
 \author{Paul D. Lasky}
 \author{Daniel J. Price}
\affiliation{Monash Centre for Astrophysics and School of Physics and Astronomy, Monash University, VIC 3800, Australia}


\date{\today}

\begin{abstract}
 We perform three-dimensional numerical relativity simulations of homogeneous and inhomogeneous expanding spacetimes, with a view towards quantifying non-linear effects from cosmological inhomogeneities. We demonstrate fourth-order convergence with errors less than one part in $10^{6}$ in evolving a flat, dust Friedmann-Lema\^itre-Roberston-Walker (FLRW) spacetime using the Einstein Toolkit within the Cactus framework. We also demonstrate agreement to within one part in $10^{3}$ between the numerical relativity solution and the linear solution for density, velocity and metric perturbations in the Hubble flow over a factor of $\sim350$ change in scale factor (redshift). We simulate the growth of linear perturbations into the non-linear regime, where effects such as gravitational slip and tensor perturbations appear. We therefore show that numerical relativity is a viable tool for investigating nonlinear effects in cosmology.


\end{abstract}

\pacs{Valid PACS appear here}
\maketitle

\section{\label{sec:intro}Introduction}
 Modern cosmology relies on the cosmological principle --- that the Universe is sufficiently homogeneous and isotropic on large scales to be described by a Friedmann-Lema\^itre-Robertson-Walker (FLRW) model. Cosmological N-body simulations \citep[e.g.][]{genel2014,springel2005,kim2011} encode these assumptions by prescribing the expansion to be that of the FLRW model, governed by the Friedmann equations, while employing a Newtonian approximation for gravity.
 

The transition to cosmic homogeneity begins on scales $\sim 80 h^{-1}$ Mpc \cite[e.g.][]{yadav2010,scrimgeour2012}, but is inhomogeneous and anisotropic on smaller scales.
Upcoming cosmological surveys utilising Euclid, the Square Kilometre Array (SKA) and the Large Synoptic Survey Telescope (LSST)  \citep{amendola2016,maartens2015,ivezic2008} will reach a precision at which nonlinear general relativistic effects from these inhomogeneities could be important. A more extreme hypothesis \citep{rasanen2004,kolb2005,kolb2006, notari2006, rasanen2006a,rasanen2006b,li2007,li2008, larena2009,buchert2015,green2016,bolejkolasky2008} is that such inhomogeneities may provide an alternative explanation for the accelerating expansion of the Universe, via \emph{backreaction} \citep[see][for a review]{buchert2008,buchert2012}, replacing the role assigned to dark energy in the standard $\Lambda$CDM model \citep{riess1998,perlmutter1999,parkinson2012,samushia2013}. 

Quantifying the general relativistic effects associated with nonlinear structures ultimately requires solving Einstein's equations. Post-Newtonian approximations are a worthwhile approach \cite{matarrese1996,rasanen2010,green2011,green2012,adamek2013,adamek2016a,adamek2016b,sanghai2015,oliynyk2014,noh2004}, however the validity of these must be checked against a more precise solution since the density perturbations themselves are highly nonlinear.

 An alternative approach is to use numerical relativity, which has enjoyed tremendous success over the past decade \citep{pretorius2005,campanelli2006,baker2006}.
Cosmological modelling with numerical relativity began with evolutions of planar and spherically symmetric spacetimes using the Arnowitt-Deser-Misner (ADM) formalism \citep{arnowitt1959}, including Kasner and matter-filled spacetimes \citep{centrella1979}, the propagation and collision of gravitational wave perturbations \citep{centrella1980, centrella1982} and linearised perturbations to a homogeneous spacetime \citep{centrella1983,centrella1984}. More recent work has continued to include symmetries to simplify the numerical calculations \citep[e.g.][]{rekier2015,torres2014}.

 

Simulations free of these symmetries have only emerged within the last year. \citet{giblin2016a} studied the evolution of small perturbations to an FLRW spacetime, exploring observational implications in \citep{giblin2016b}. \citet{bentivegna2015} showed differential expansion in an inhomogeneous universe, and quantified the backreaction parameter from \citep{buchert2000a} for a single mode perturbation. These works all indicate that the effects of nonlinear inhomogeneities may be significant.

In this work, we perform a feasibility study of numerical solutions to the full Einstein equations for inhomogeneous cosmologies by simulating the growth of structure in a model three-dimensional universe and comparing to known analytic solutions. Our approach is similar to \citep{giblin2016a,bentivegna2015,giblin2016b}, with differences in the generation of initial conditions and numerical methods. We use the freely-available Einstein Toolkit, based on the Cactus infrastructure \citep{loffler2012,zilhao2013}. We benchmark our three-dimensional numerical implementation on two analytic solutions of Einstein's equations relevant to cosmology: FLRW spacetime and the growth of linear perturbations. We also present the growth of perturbations into the nonlinear regime, and analyse the resulting gravitational slip \citep{daniel2008,daniel2009} and tensor perturbations. 

In Section \ref{sec:method} we describe our numerical methods, including gauge choices (\ref{sec:gauge}) and an overview of the derivations of the linearly perturbed Einstein equations used for our initial conditions (\ref{sec:initial_conditions}). In Section \ref{sec:FLRW} we describe the setup (\ref{sec:FLRW_setup}) and results (\ref{sec:FLRW_results}) of our evolutions of a flat, dust FLRW universe. The derivation of initial conditions for linear perturbations to the FLRW model are described in \ref{sec:linear_setup}, with results presented in \ref{sec:linear_results}. The growth of the perturbations to nonlinear amplitude is presented in \ref{sec:nonlinear}, with analysis of results and higher order effects in \ref{sec:nonlinear_results} and \ref{sec:nonlinear_highorder} respectively. We adopt geometric units with $G=c=1$, Greek indices run from 0 to 3 while Latin indices run from 1 to 3, with repeated indices implying summation.

\section{\label{sec:method}Numerical Method}
We integrate Einstein's equations with the Einstein Toolkit, a free, open-source code for numerical relativity \citep{loffler2012}. This utilises the Cactus infrastructure, consisting of a central core, or "flesh", with application modules called "thorns" that communicate with this flesh \citep{cactus}. The Einstein Toolkit is a collection of thorns for computational relativity, used extensively for simulations of binary neutron star and black hole mergers \citep[e.g][]{kastaun2015,radice2015,baiotti2005}. Numerical cosmology with the Einstein Toolkit is a new field \citep{bentivegna2015}. We use the \texttt{McLachlan} code \citep{brown2009} to evolve spacetime using the Baumgarte-Shapiro-Shibata-Nakamura (BSSN) formalism \citep{shibata1995, baumgarte1999}, and the \texttt{GRHydro} code to evolve the hydrodynamical system \citep{baiotti2005, giacomazzo2007, mosta2014}; a new setup for cosmology with the Einstein Toolkit.

We use the fourth-order Runge-Kutta method, adopt the Marquina Riemann solver and use the piecewise parabolic method for reconstruction on cell interfaces. \texttt{GRHydro} is globally second order in space due to the coupling of hydrodynamics to the spacetime \citep{hawke2005, mosta2014}. We therefore expect fourth-order convergence of our numerical solutions for the spatially homogeneous FLRW model. Once perturbations are introduced to this model we expect our solutions to be second-order accurate. 

We have developed a new thorn, \texttt{FLRWSolver}, to initialise an FLRW cosmological setup with optional linear perturbations. We evolve our simulations in a cubic domain on a uniform grid with periodic boundary conditions with $x^{i}$ in [-240,240]. Our domain sizes are $20^{3}, 40^{3}$ and $80^{3}$, respectively using 70 (8 cores), 380 (8 cores) and 790 (16 cores) CPU hours. 

\subsection{\label{sec:gauge}Gauge}
The gauge choice corresponds to a choice of the lapse function, $\alpha$, and shift vector, $\beta^{i}$. The metric written in the $(3+1)$ formalism is
\begin{equation}\label{eq:3p1_metric}
	ds^{2} = -\alpha^{2}dt^{2} + \gamma_{ij}(dx^{i} + \beta^{i}dt)(dx^{j} + \beta^{j}dt), 
\end{equation}
where $\gamma_{ij}$ is the spatial metric. Previous cosmological simulations with numerical relativity adopt the synchronous gauge, corresponding to $\alpha=1, \;\beta^{i}=0$ \citep{giblin2016a, bentivegna2015}. We instead utilise the general spacetime foliation of \citep{bona1995},
\begin{equation}
	\partial_{t}\alpha = -\alpha^{2} \, f(\alpha) \, K, \label{eq:harmonic_gauge}
\end{equation}
where $f(\alpha)>0$ is an arbitrary function, and $K=\gamma^{ij}K_{ij}$. We set the shift vector $\beta^{i}=0$.
Harmonic slicing uses $f$ = const., while $f=1/\alpha$ corresponds to the "1+log" slicing common in black hole binary simulations. We choose harmonic slicing with $f=0.25$ to maintain the stability of our evolutions, as in \citep{torres2014}. Harmonic slicing also allows for longer evolutions for the same computational time, compared to 1+log slicing, due to the increased rate of change of the lapse. We adopt this gauge for numerical convenience, and acknowledge possible alternative methods include using synchronous gauge with adaptive time-stepping.
We use \eqref{eq:harmonic_gauge} for evolution only. We scale to the gauge described in the next section for analysis.


\subsection{\label{sec:initial_conditions}Perturbative Initial Conditions}
Bardeen's formalism of cosmological perturbations \citep{bardeen1980} was developed with the intention to connect metric perturbations to physical perturbations in the Universe. This connection is made clear by defining the perturbations as gauge-invariant quantities in the longitudinal gauge.
The general line element of a perturbed, flat FLRW universe, including scalar ($\Phi, \Psi$), vector ($B_{i}$) and tensor ($h_{ij}$) perturbations takes the form
\begin{equation}\label{eq:perturbed_metric_svt}
	\begin{aligned}
		ds^{2} = a^{2}(\eta)[-(1 &+ 2\Psi)d\eta^{2} - 2B_{i}dx^{i}d\eta \\
		&+ (1-2\Phi)\delta_{ij} dx^{i}\,dx^{j}+ h_{ij}dx^{i}\,dx^{j}], 
	\end{aligned}
\end{equation}
where $\eta$ is conformal time, $a(\eta)$ is the FLRW scale factor and $\delta_{ij}$ is the identity matrix. 
We derive initial conditions from the linearly perturbed Einstein equations, implying negligible vector and tensor perturbations \citep{adamek2013}. This is valid as long as our simulations begin at sufficiently high redshift that the Universe may be approximated by an FLRW model with small perturbations. 
Considering only scalar perturbations the metric becomes
\begin{equation}
	ds^{2} = a^{2}(\eta)[-(1+2\Psi)d\eta^{2} + (1-2\Phi)\delta_{ij}dx^{i}dx^{j}], \label{eq:perturbed_metric}
\end{equation}
where $\Phi$ and $\Psi$ are Bardeen's gauge-invariant scalar potentials \citep{bardeen1980}. Here we see that $\Psi$, the Newtonian potential, will largely influence the motion of non-relativistic particles; where the time-time component of the metric dominates the motion. The Newtonian potential plays the dominant role in galaxy clustering. Relativistic particles will \emph{also} be affected by the curvature potential $\Phi$, and so both potentials influence effects such as gravitational lensing \citep{bertschinger2011, bardeen1980}.

The metric perturbations are coupled to perturbations in the matter distribution via the stress-energy tensor. We approximate the homogeneous and isotropic background as a perfect fluid in thermodynamic equilibrium, giving
\begin{equation}
	T_{\mu\nu} = \left(\rho + P\right)u_{\mu}u_{\nu} + P\,g_{\mu\nu}, \label{eq:stress_energy_full}
\end{equation}
where $\rho$ is the total energy density, $P$ is the pressure and $u^{\mu}$ is the four-velocity of the fluid. 
We assume a dust universe, implying negligible pressure ($P\ll\rho$), and we solve the perturbed Einstein equations,
\begin{equation}
	\delta G_{\mu\nu} = 8\pi \, \delta T_{\mu\nu}, \label{eq:perturbed_einstein}
\end{equation}
using linear perturbation theory. From the time-time, time-space, trace and trace free components of \eqref{eq:perturbed_einstein}, we obtain the following system of equations \citep{sachs1967,adamek2013}
\begin{subequations} \label{eqs:perturbed_einstein}
	\begin{align}
		\nabla^{2}\Phi - 3H\left(\dot{\Phi} + H \Psi\right) &= 4\pi  \bar{\rho}\,\delta a^{2}, \label{eq:einstein_1} \\ 
		H \partial_{i}\Psi + \partial_{i}\dot{\Phi} &= -4\pi \bar{\rho} \,a^{2} \delta_{ij}\delta v^{j}, \label{eq:einstein_2} \\ 
		\ddot{\Phi} + H\left(\dot{\Psi} + 2\dot{\Phi}\right) &= \frac{1}{2}\nabla^{2}(\Phi - \Psi), \label{eq:einstein_3} \\ 
		\partial_{\langle i}\partial_{j\rangle} \left(\Phi - \Psi\right) &= 0. \label{eq:einstein_4}
	\end{align}	
\end{subequations}
Here $H\equiv \dot{a}/a$ is the Hubble parameter, $\partial_{i} \equiv \partial / \partial x_{i}$, $\nabla^{2}=\partial^{i}\partial_{i}$, $\partial_{\langle i}\partial_{j\rangle}\equiv \partial_{i}\partial_{j} - 1/3\,\delta_{ij}\nabla^{2}$, and a dot represents a derivative with respect to conformal time $\eta$. The quantity $|\Phi-\Psi|$ is known as the gravitational slip \citep{daniel2008,daniel2009,bertschinger2011}, which is zero in the linear regime and in the absence of anisotropic stress. At higher orders in perturbation theory, the gravitational slip is non-zero, and $\Phi\neq\Psi$ \citep[see e.g.][]{ballesteros2012}.

We perturb the density and coordinate three-velocity by making the substitutions
\begin{subequations} \label{eqs:matter_perturb}
	\begin{align}
		\rho &= \bar{\rho}\,(1+\delta), \\ 
		v^{i} &= \delta v^{i},
	\end{align}
\end{subequations}
where $\bar{\rho}$ represents the background FLRW density, and $\bar{v}^{i}=0$.
We derive the relativistic fluid equations from the components of the energy-momentum conservation law,
\begin{equation}
	\nabla_{\alpha} T_{\mu}^{\phantom{\alpha}\alpha} = 0,
\end{equation}
where $\nabla_{\alpha}$ is the covariant derivative associated with the 4-metric. The resulting continuity and Euler equations are, 
\begin{subequations}
	\begin{align}
		\dot{\delta} &= 3\dot{\Phi} - \partial_{i}v^{i}, \label{eq:rel_continuity} \\
		Hv^{i} + \dot{v}^{i} &= -\partial^{i}\Psi. \label{eq:rel_euler}
	\end{align}
\end{subequations}

\begin{figure*}[!ht]
	\includegraphics[width=\textwidth]{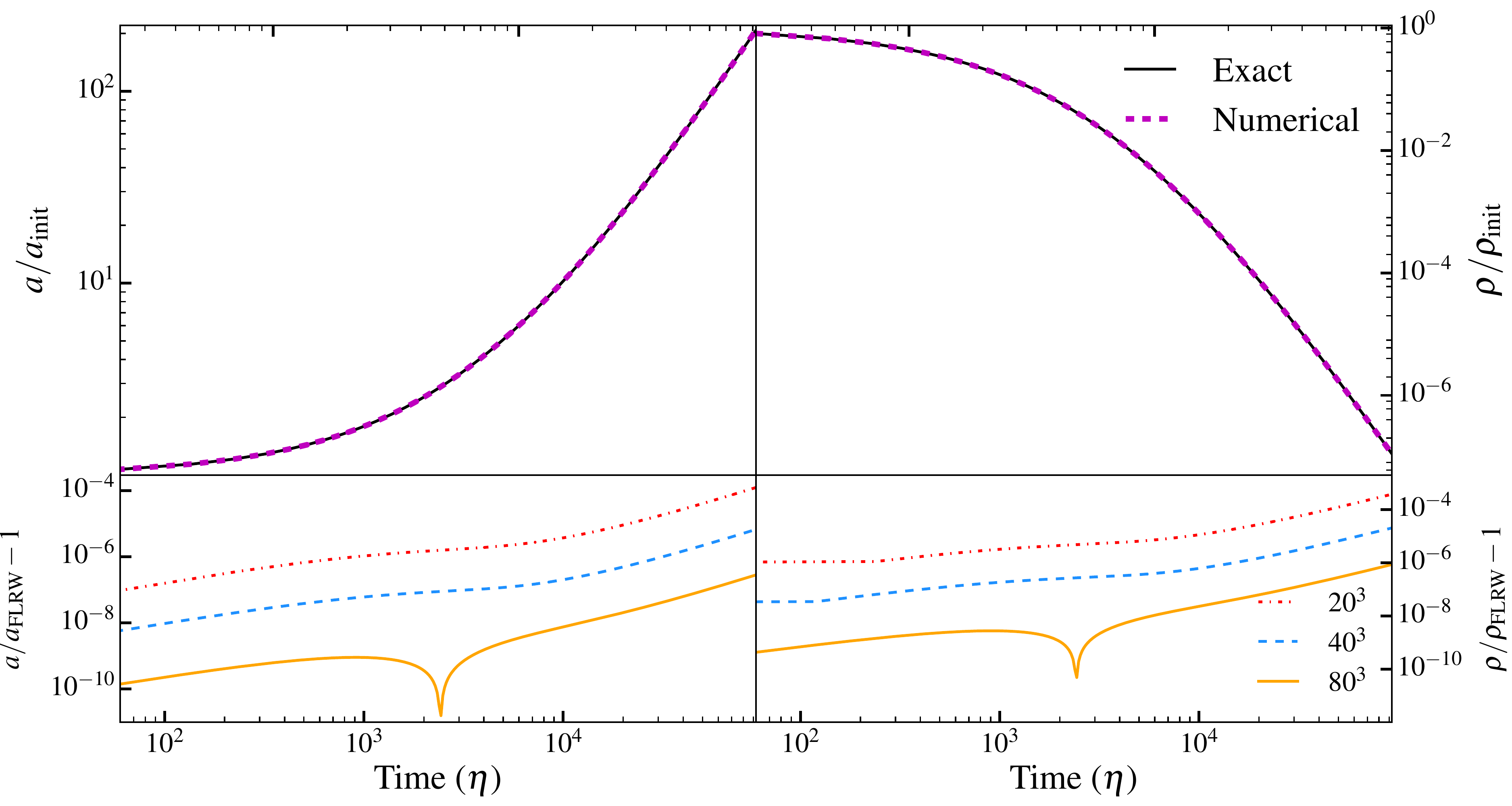}
	\caption{\label{fig:FLRW_a_rho} Comparison between our numerical simulations (magenta) and the exact solutions (black) for a dust FLRW universe. Top: evolution of the scale factor, $a$ (left) and the density, $\rho$ (right), relative to their initial values $a_{\mathrm{init}}$ and $\rho_{\mathrm{init}}$, as a function of conformal time $\eta$. Bottom: errors in the FLRW scale factor (left) and density (right) at domain sizes $20^{3}, \,40^{3}$ and $80^{3}$.}
\end{figure*}
\begin{figure*}[!ht]
	\includegraphics[width=\textwidth]{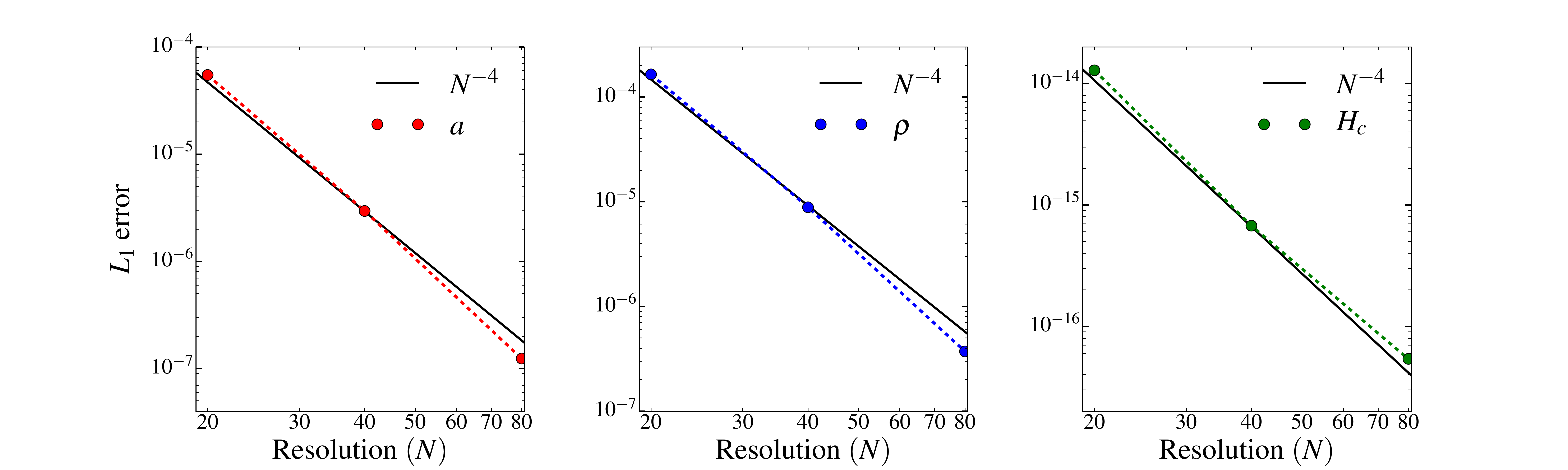}
	\caption{\label{fig:FLRW_RMS}Fourth-order convergence in the FLRW calculations, showing $L_{1}$ error as a function of resolution for the scale factor (left), density (middle), and Hamiltonian constraint (right). $N$ refers to the number of grid points along one spatial dimension. Filled circles indicate data points from our simulations, dashed lines join these points, and black solid lines indicate the expected $N^{-4}$ convergence.}
\end{figure*}


\section{\label{sec:FLRW}FLRW spacetime}
We test our thorn \texttt{FLRWSolver} together with the Einstein toolkit on two analytic solutions to Einstein's equations relevant to cosmology. Our first and simplest test is the flat, dust FLRW model. Here we initialise a homogeneous and isotropic matter distribution and spatial metric, and evolve in the harmonic gauge, as outlined in section \ref{sec:gauge}. While the Einstein Toolkit has been previously tested on FLRW and Kasner cosmologies \citep{loffler2012, vulcanov2002}, this is an important first test of \texttt{FLRWSolver} and its interaction with the evolution thorns. 

\subsection{\label{sec:FLRW_setup}Setup}
The line element for a spatially homogeneous and isotropic FLRW spacetime is given by
\begin{equation}
	ds^{2} = a^{2}(\eta)[ - d\eta^{2} + \frac{1}{(1+kr^{2}/4)^{2}}\delta_{ij}dx^{i}dx^{j}],
\end{equation}
where $k=-1,0,1$ if the universe is open, flat or closed respectively.
Assuming homogeneity and isotropy Einstein's equations reduce to the Friedmann equations \citep{friedmann1922, friedmann1924}, 
\begin{subequations}\label{eq:Friedmann}
	\begin{align}
		\left(\frac{\dot{a}}{a}\right)^{2} &= \frac{8\pi\rho \,a^{2}}{3} - k, \label{eq:friedmann_1}\\
		\dot{\rho} &= -3\frac{\dot{a}}{a}\left(\rho + P\right). \label{eq:friedmann_2}
	\end{align}
\end{subequations}
In the remainder of the paper we assume a flat spatial geometry, supported by combined Planck and Baryon Acoustic Oscillation data \citep{planck2015params}. The flat ($k=0$), dust ($P\ll\rho$) solution to \eqref{eq:Friedmann} is
\begin{equation}
	\frac{a}{a_{\mathrm{init}}} = \xi^{2}, \quad
	\frac{\rho}{\rho_{\mathrm{init}}} = \xi^{-6},
\end{equation}
where $a_{\mathrm{init}}, \rho_{\mathrm{init}}$ are the values of $a, \rho$ at $\eta=0$ respectively, and we have introduced the scaled conformal time coordinate
\begin{equation}
	\xi\equiv 1 + \sqrt{\frac{2\pi\rho^{*}}{3a_{\mathrm{init}}}}\,\eta,
\end{equation}
where $\rho^{*}=\rho\,a^{3}$ is the conserved (constant) comoving density for an FLRW universe. The familiar $\tau^{2/3}$ solution for the scale factor arises in the Newtonian gauge with $ds^{2} = -d\tau^{2} + \gamma_{ij}dx^{i}dx^{j}$ (for a flat spacetime; see Appendix \ref{appx:newt_gauge}).

We initialise a homogeneous and isotropic matter distribution by specifying constant density $\rho_{\mathrm{init}}=10^{-8}$ and zero velocity in \texttt{FLRWSolver}, with $a_{\mathrm{init}}=1$. The Einstein Toolkit then initialises the stress-energy tensor, coupled to our homogeneous and isotropic spacetime, characterised by the spatial metric, $\gamma_{ij} = a^{2}(\eta)\delta_{ij}$, and extrinsic curvature, also set in \texttt{FLRWSolver}. We define the extrinsic curvature via the relation
\begin{equation}
 	\frac{d}{dt}\gamma_{ij} = -2\alpha K_{ij}, \label{eq:excurv_def}
\end{equation}	
where $d/dt = \partial / \partial t - \mathcal{L}_{\beta}$, and $\mathcal{L}_{\beta}$ is the Lie derivative with respect to the shift vector. Since we choose $\beta^{i}=0$, we have $d/dt = \partial / \partial t$. The extrinsic curvature for our FLRW setup is therefore
\begin{equation}
	K_{ij} = -\frac{\dot{a}a}{\alpha}\delta_{ij}.
\end{equation}
We evolve the system in the harmonic gauge until the domain volume has increased by one million, corresponding to a change in redshift of $\sim100$. 

To analyse our results we scale the time from the metric \eqref{eq:3p1_metric} to the longitudinal gauge \eqref{eq:perturbed_metric} using the coordinate transform $t=t(\eta)$. 
This gives
\begin{equation}
	\frac{dt}{d\eta} = \frac{a(\eta)}{\alpha(t)}, 
\end{equation}
which we integrate to find the scaled conformal time in terms of $t$ to be
\begin{equation}
	\xi(t) = \left(\sqrt{6\pi\rho_{\mathrm{init}}} \,\int\alpha(t)\,dt + 1\right)^{1/3},
\end{equation}
where we numerically integrate the lapse function $\alpha$ using the trapezoidal rule. This coordinate transformation allows us to simulate longer evolutions for less computational time, while still performing our analysis in the longitudinal gauge to extract physically meaningful results.




\subsection{\label{sec:FLRW_results}Results}
Figure~\ref{fig:FLRW_a_rho} compares our numerical relativity solutions with the exact solutions to the Friedmann equations. The top panels show the time evolution of $a$ and $\rho$ (dashed magenta curves) relative to their initial values, which may be compared to the exact solutions, $a_{\mathrm{FLRW}}$ and $\rho_{\mathrm{FLRW}}$ (black solid curves). The bottom panels show the residuals in our numerical solutions at resolutions of $20^{3}$, $40^{3}$ and $80^{3}$. The error can be seen to decrease when the spatial resolution is increased. The increase in spatial resolution causes the timestep to decrease via the Courant condition. To quantify this, we compute the $L_{1}$ error, given by (e.g. for the scale factor)
\begin{equation}
	L_{1}(a) = \frac{1}{n} \sum_{i=1}^{n} \left|\frac{a}{a_{FLRW}}-1\right|,
\end{equation}
where $n$ is the total number of timesteps.
As outlined in Section \ref{sec:method}, we expect fourth-order convergence due to the spatial homogeneity. Figure~\ref{fig:FLRW_RMS} demonstrates this is true for the scale factor (left), density (middle) and the Hamiltonian constraint (right), 
\begin{equation}
	H \equiv \,^{(3)}R - K_{ij}K^{ij} + K^{2} -16\pi\rho = 0, \label{eq:Hamiltonian}
\end{equation}
where $^{(3)}$R is the 3-Riemann scalar and $K=\gamma^{ij}K_{ij}$. For the FLRW model this reduces to the first Friedmann equation \eqref{eq:friedmann_1}.

The results of this test demonstrate that the Einstein Toolkit, in conjunction with our initial-condition thorn \texttt{FLRWSolver}, produces agreement with the exact solution for a flat, dust FLRW spacetime, with relative errors less than $10^{-6}$, even at low spatial resolution ($80^{3}$). 


\begin{figure*}[!ht]
	\includegraphics[width=\textwidth]{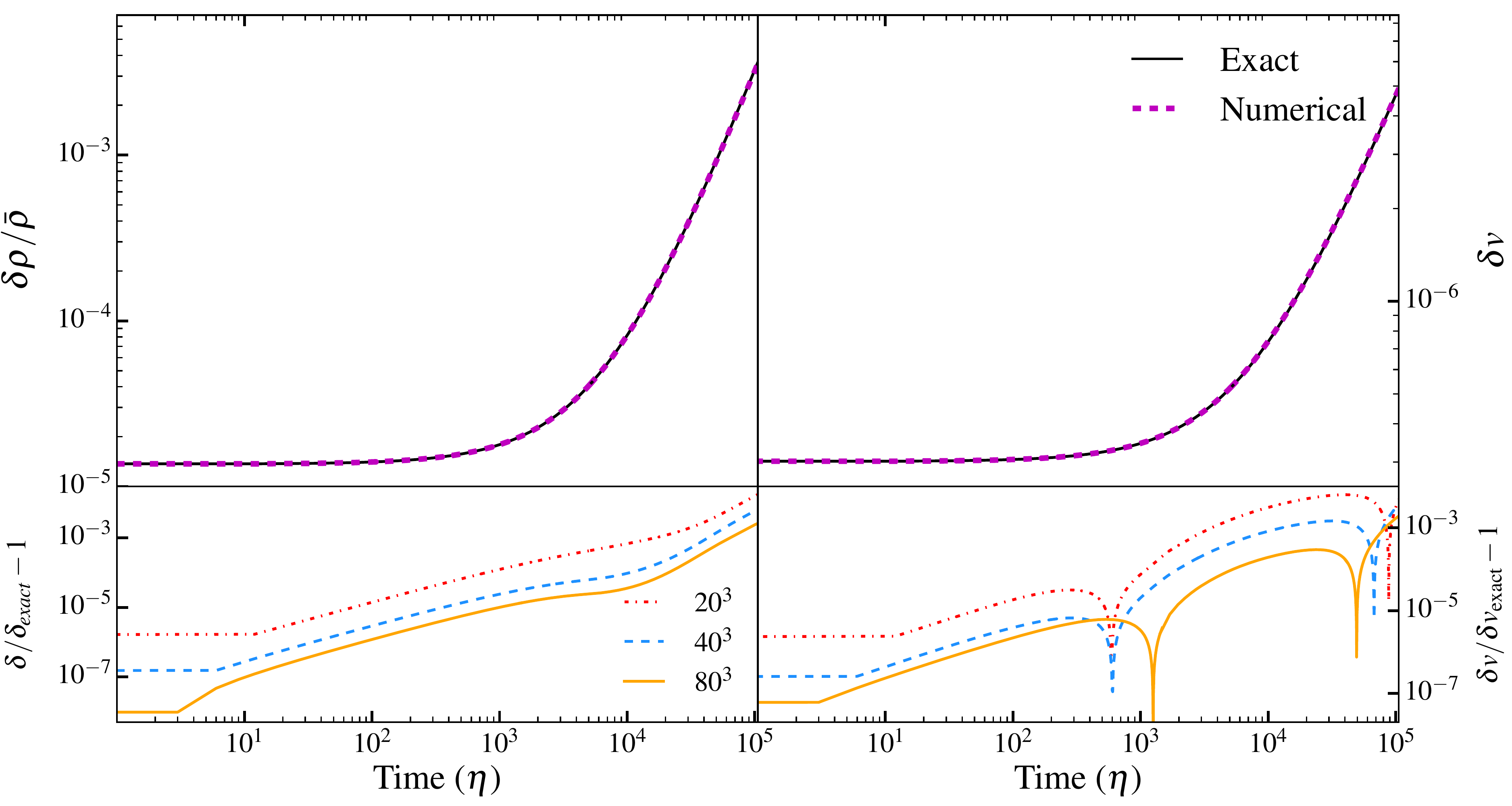}
	\caption{\label{fig:perturb_deltas}Comparison between our numerical relativity solutions and exact solutions for the linear perturbations to a dust FLRW model. We show the conformal time ($\eta$) evolution of the fractional density perturbation (top left) and the velocity perturbation (top right) computed from one-dimensional slices along the $x$ axis of our domain. Bottom: relative errors for calculations at $20^{3}$, $40^{3}$ and $80^{3}$. }
\end{figure*}
\begin{figure*}[!ht]
	\includegraphics[width=0.75\textwidth]{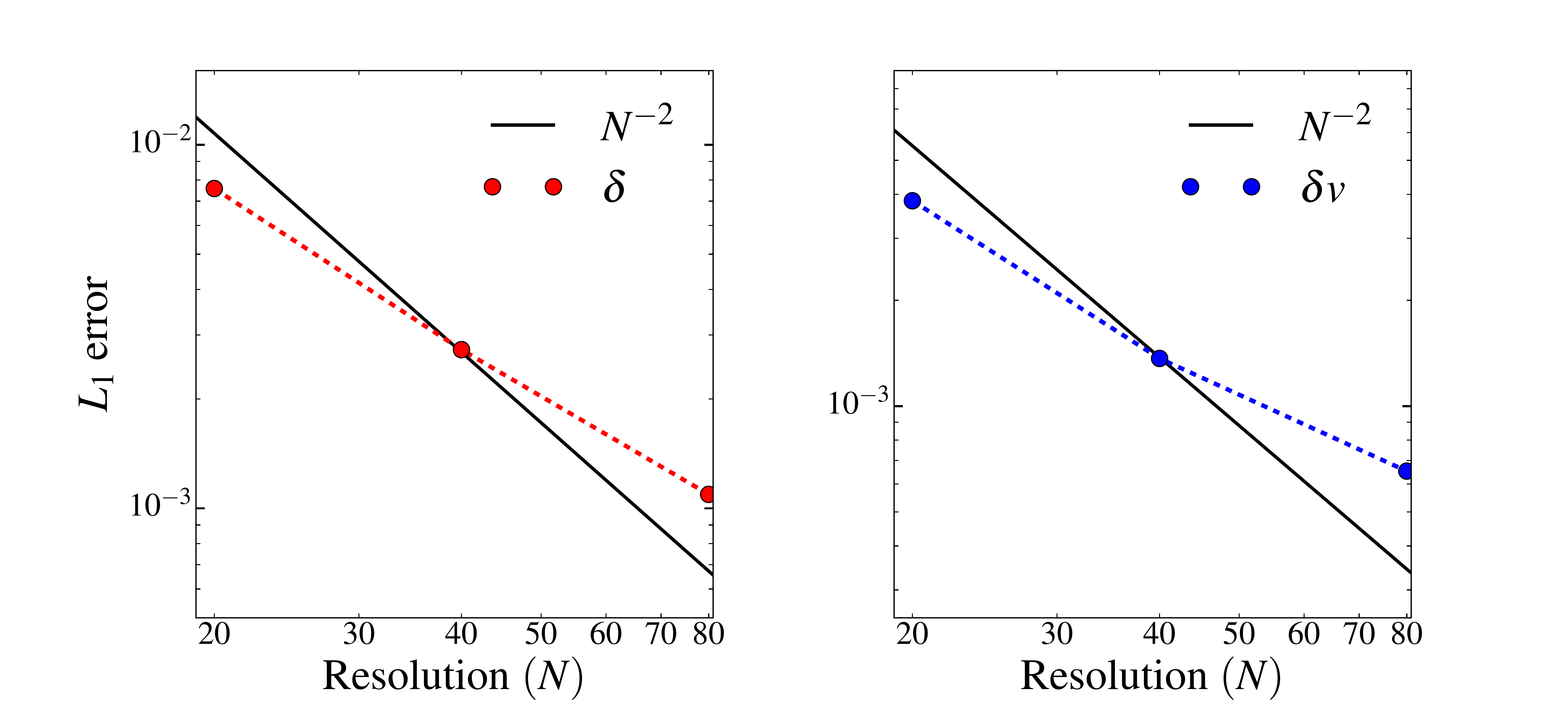}
	\caption{\label{fig:perturb_converge}Second order convergence of our numerical solutions to the exact solutions for a linearly perturbed FLRW spacetime, showing $L_{1}$ errors in the density (left) and velocity perturbations (right). $N$ refers to the number of grid points along one spatial dimension. Filled circles indicate data points from our simulations, dashed lines join these points, and black solid lines indicate the expected $N^{-2}$ convergence.}
\end{figure*}

\section{\label{sec:linear}Linear Perturbations}
For our second test we introduce small perturbations to the FLRW model. The evolution of these perturbations in the linear regime can be found by solving the system of equations \eqref{eqs:perturbed_einstein}. We use these solutions (derived below) to set the initial conditions.

\subsection{\label{sec:linear_setup}Setup}
In the absence of anisotropic stress we have $\Psi=\Phi$. Equation \eqref{eq:einstein_3} then becomes purely a function of $\Phi$ and the FLRW scale factor $a$. Solving this gives
\begin{equation}\label{eq:phi_full}
	\Phi = f(x^{i}) - \frac{g(x^{i})}{5\,\xi^{5}}, 
\end{equation}
where $f, g$ are functions of only the spatial coordinates. We substitute \eqref{eq:phi_full} into the Hamiltonian constraint, Equation \eqref{eq:einstein_1}, to give the fractional density perturbation $\delta \equiv \delta\rho / \bar{\rho}$, in the form
\begin{equation}\label{eq:delta_full}
	\begin{aligned} 
		\delta = C_{1}\, \xi^{2}\, \nabla^{2}&f(x^{i}) - 2 \,f(x^{i}) \\
		&- C_{2} \,\xi^{-3}\,\nabla^{2}g(x^{i}) - \frac{3}{5} \xi^{-5} g(x^{i}),
	\end{aligned}
\end{equation}
where we have defined
\begin{equation}
	C_{1}\equiv \frac{a_{\mathrm{init}}}{4\pi\rho^{*}},\quad C_{2}\equiv \frac{a_{\mathrm{init}}}{20\pi\rho^{*}}. \label{eq:c1c2}
\end{equation}
Using the momentum constraint, Equation \eqref{eq:einstein_2}, the velocity perturbation $\delta v^{i}$ is therefore
\begin{equation}\label{eq:vel_full}
	\delta v^{i} = C_{3}\,\xi\, \partial^{i}f(x^{i}) + \frac{3}{10}C_{3}\,\xi^{-4}\, \partial^{i}g(x^{i})
\end{equation}
where we have
\begin{equation}
	C_{3}\equiv-\sqrt{\frac{a_{\mathrm{init}}}{6\pi\rho^{*}}}. \label{eq:c3c4}
\end{equation}
Equation~\eqref{eq:delta_full} demonstrates both a growing and decaying mode for the density perturbation \citep{bardeen1980, mukhanov1992}. We set $g(x^{i})=0$ to extract only the growing mode, giving
\begin{subequations} \label{eqs:exact_solns_growing}	
	\begin{align}
		\Phi &= f(x^{i}), \label{eq:phi} \\
		\delta &= C_{1}\, \xi^{2}\, \nabla^{2}f(x^{i}) - 2 \,f(x^{i}), \label{eq:delta}\\
		\delta v^{i} &= C_{3}\,\xi\, \partial^{i}f(x^{i}), \label{eq:vel}
	\end{align}
\end{subequations}
from which we set our initial conditions. 
We choose
\begin{equation}
	\Phi = \Phi_{0} \sum_{i=1}^{3} \mathrm{sin}\left(\frac{2\pi x^{i}}{L}\right),
\end{equation}
where $L$ is the length of one side of our computational domain. We require the amplitude $\Phi_{0}\ll1$ so that our assumptions of linearity are valid, and so we set $\Phi_{0}=10^{-8}$.
This choice then sets the form of our density and velocity perturbations, as per \eqref{eq:delta} and \eqref{eq:vel}. At $\eta=0$ ($\xi=1$) these are,
\begin{align}
	\delta &= \left[ \left(\frac{2\pi}{L}\right)^{2}C_{1} - 2\right] \Phi_{0} \sum_{i=1}^{3} \mathrm{sin}\left(\frac{2\pi x^{i}}{L}\right),\label{eq:initial_delta}\\
	\delta v^{i} &= \frac{2\pi}{L}C_{3}\, \Phi_{0} \sum_{i=1}^{3} \mathrm{cos}\left(\frac{2\pi x^{i}}{L}\right), \label{eq:initial_deltav}
\end{align}
and the choice of $\Phi_{0}$ results in amplitudes of $\delta\sim 10^{-5}$ and $\delta v^{i} \sim 10^{-7}$.
We set these matter perturbations in \texttt{FLRWSolver}, implementing negligible pressure and again using \eqref{eq:excurv_def} to define the extrinsic curvature. For a linearly perturbed FLRW spacetime with $\Psi=\Phi$ and $\dot{\Phi}=0$ we have
\begin{equation}
	K_{ij} = -\frac{\dot{a}a}{\alpha}(1-2\Phi)\delta_{ij}.
\end{equation}
We evolve these perturbations in the harmonic gauge until the volume of the domain has increased by 125 million, $(\Delta a)^{3} \sim 1.25\times10^{8}$, corresponding to a factor of 500 change in redshift. 

\subsection{\label{sec:linear_results}Results}
Dashed magenta curves in Figure~\ref{fig:perturb_deltas} show the conformal time evolution of the fractional density perturbation, $\delta\equiv \delta\rho / \bar{\rho}$ (top left), and the velocity perturbation, $\delta v$ (top right). Solid black curves show the solutions \eqref{eq:delta} for $\delta_{\mathrm{exact}}$ and \eqref{eq:vel} for $\delta v_{\mathrm{exact}}$. Bottom panels show the relative errors for three different resolutions. 
Figure~\ref{fig:perturb_converge} shows the $L_{1}$ error as a function of resolution, demonstrating the expected second-order convergence.
\begin{figure}[!ht]
	\includegraphics[width=\columnwidth]{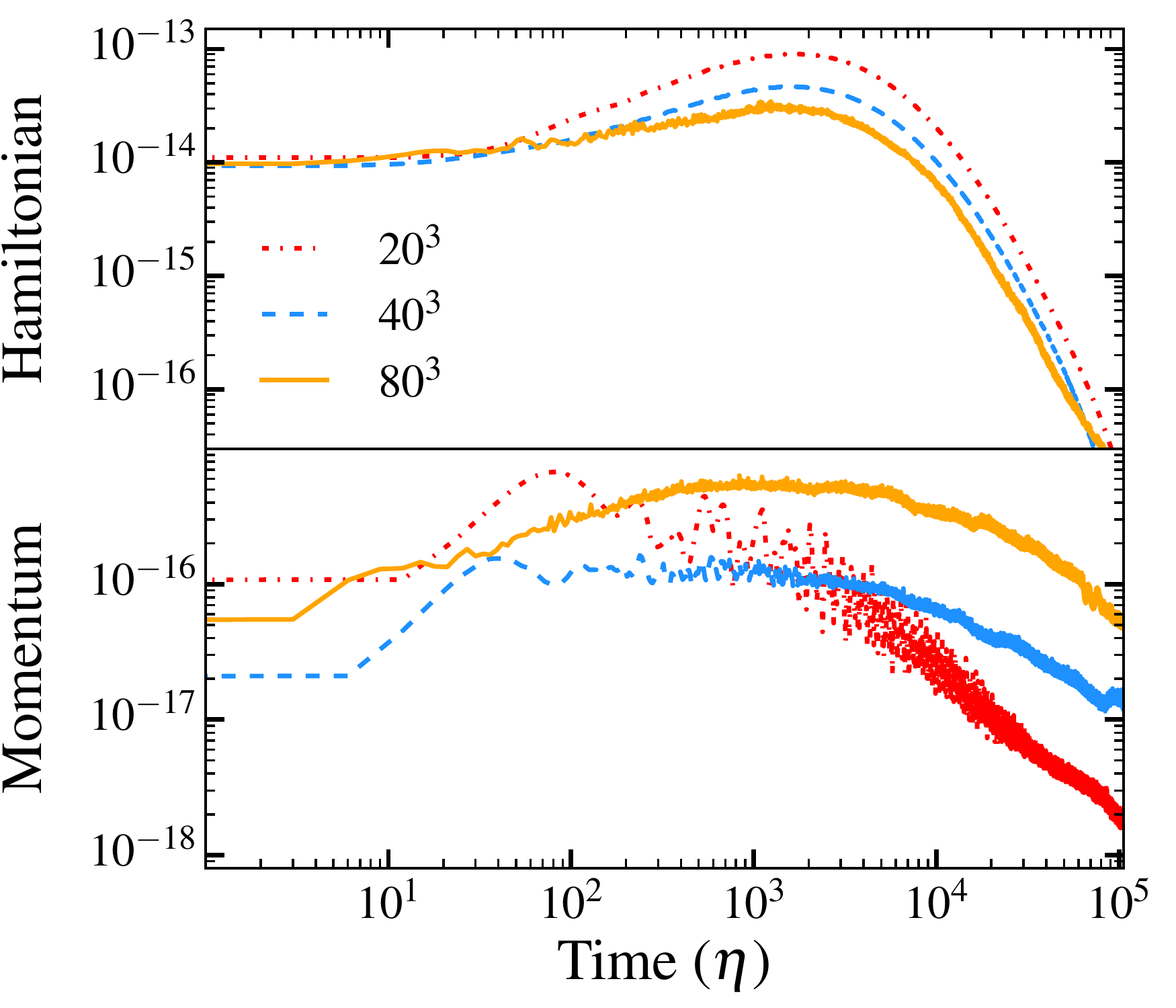}
	\caption{\label{fig:perturb_ham_mom}Maximum (anywhere in the domain) of the Hamiltonian (top panel) and momentum (bottom panel) constraints in a linearly perturbed FLRW spacetime. We show evolution over conformal time $\eta$ at resolutions $20^{3}$, $40^{3}$ and $80^{3}$.}
\end{figure}
Figure~\ref{fig:perturb_ham_mom} shows the Hamiltonian (top), and momentum (bottom) constraints as a function of conformal time at our three chosen resolutions. The Hamiltonian constraint was defined in Equation \eqref{eq:Hamiltonian}. For our linearly perturbed FLRW spacetime this reduces to Equation \eqref{eq:einstein_1}.
The momentum constraint is
\begin{equation}
	M_{i} \equiv D_{j}K^{j}_{\phantom{j}i} - D_{i}K - S_{i} = 0,
\end{equation}
where $D_{j}$ is the covariant derivative associated with the 3-metric, and the matter source $S_{i} = -\gamma_{i\alpha}n_{\beta}T^{\alpha\beta}$, with $n_{\beta}$ the normal vector \citep{baumgarte1999}.
For linear perturbations this constraint reduces to Equation \eqref{eq:einstein_2}.
Figure~\ref{fig:perturb_ham_mom} shows a better preservation of the Hamiltonian constraint with increasing resolution. The momentum constraint shows the opposite. We attribute this to the momentum constraint being preserved to of order the roundoff error, which will become larger with an increase in resolution. Even at the highest resolution the momentum constraint is preserved to within $10^{-15}$.

This second test has demonstrated a match to within $\sim10^{-3}$ of our numerical relativity solutions to the exact solutions for the linear growth of perturbations, while exhibiting the expected second-order convergence.

\begin{figure*}[!ht]
	\includegraphics[width=\textwidth]{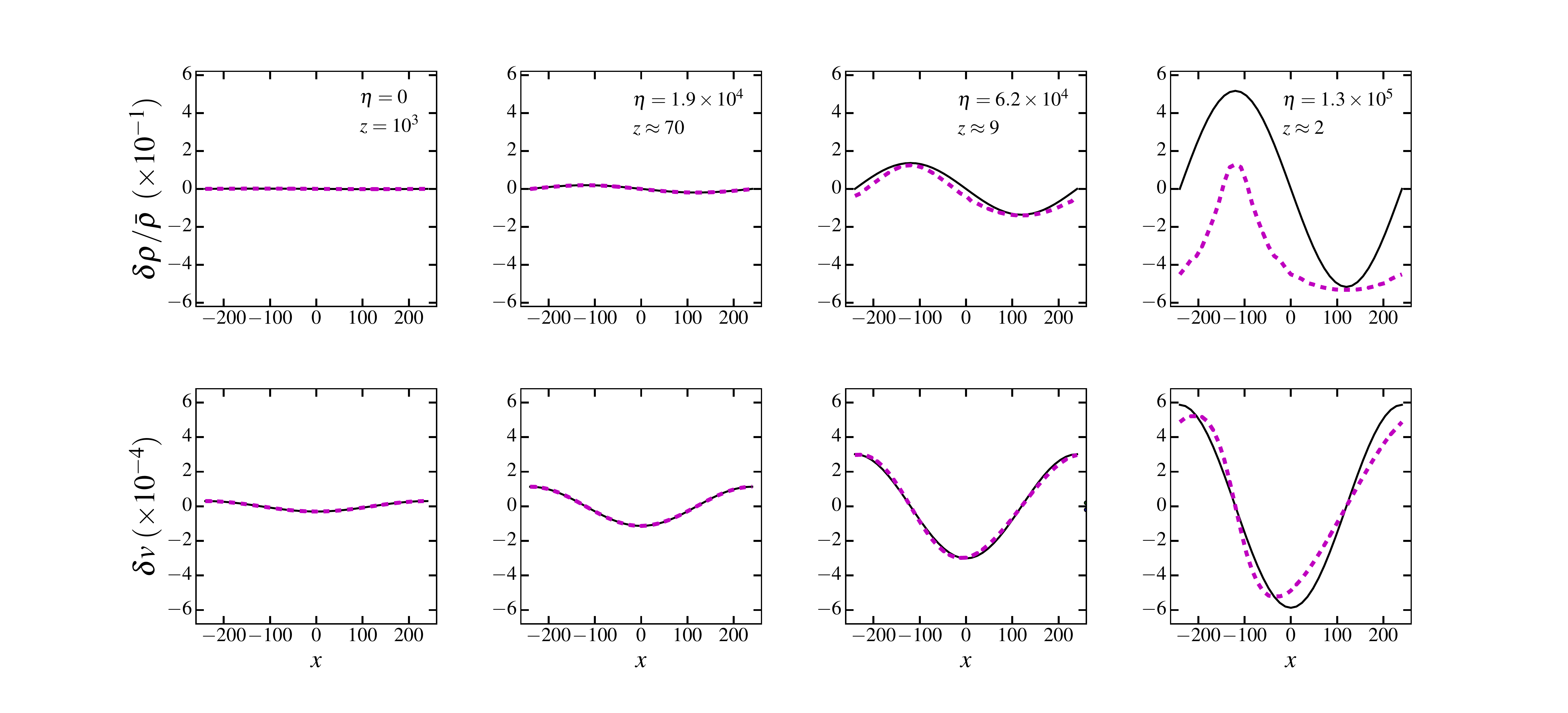} 
	\caption{\label{fig:nonlinear_x}Density (top row) and velocity (bottom row) perturbations as a function of position within our domain. Here we show one-dimensional slices of our $40^{3}$ domain during the conformal time ($\eta$) evolution. All quantities are shown in code units. Dashed magenta curves show our numerical solutions, and black solid curves show the exact solutions for the linear regime. Initial data ($\eta=0$; first column) and $\eta=1.9\times10^{4}$ (second column) match linear theory. We see a clear deviation from linear theory at $\eta=6.2\times 10^{4}$ (third column) and $\eta=1.3\times10^{5}$ (fourth column). Simulation redshifts are shown as an indicator of the \emph{change} in redshift.}
\end{figure*}
\begin{figure*}[!ht]
	\includegraphics[width=\textwidth]{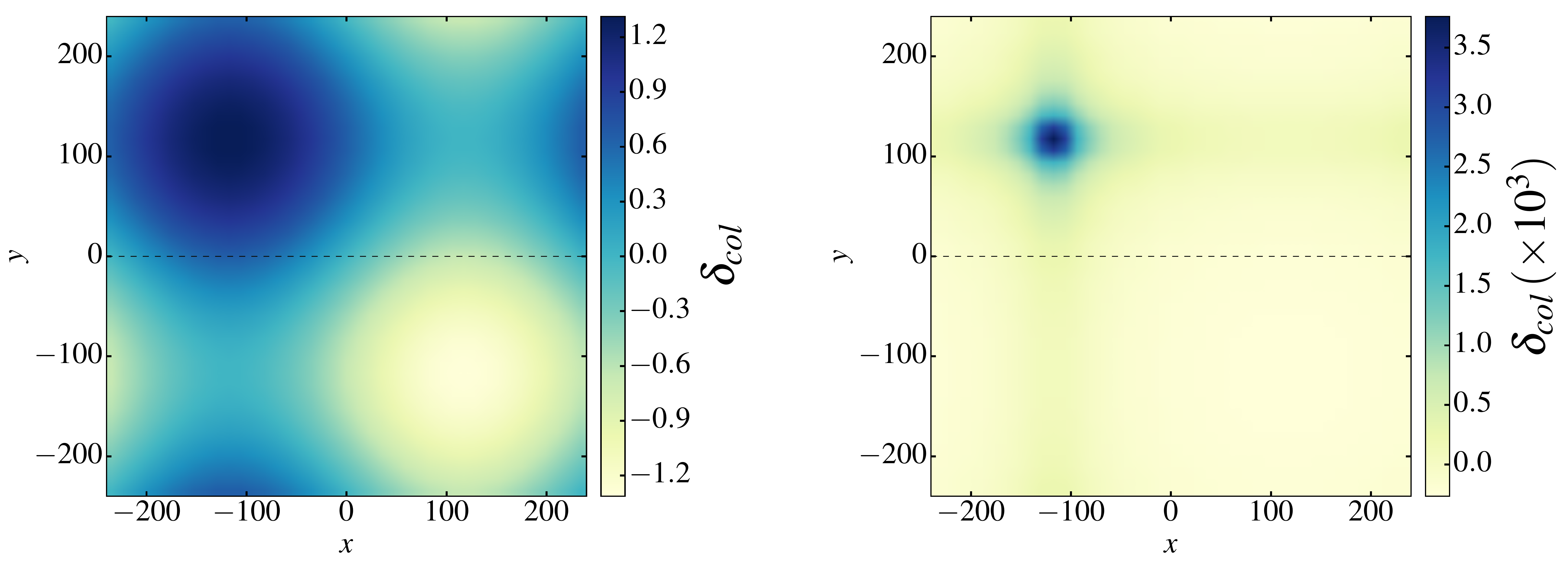}
	\caption{\label{fig:nonlinear_2D_delta} Column-integrated density perturbation showing the gravitational collapse of an overdense region. The two panels correspond to the left and right panels of Figure~\ref{fig:nonlinear_x} respectively, at conformal times of $\eta=0$ and $\eta=1.3\times10^{5}$. All quantities are shown in code units for our $40^{3}$ simulation. Grey dashed lines indicate the position of the one-dimensional slices shown in Figure~\ref{fig:nonlinear_x}.}
\end{figure*}

\begin{figure*}[!ht]
	\includegraphics[width=\textwidth]{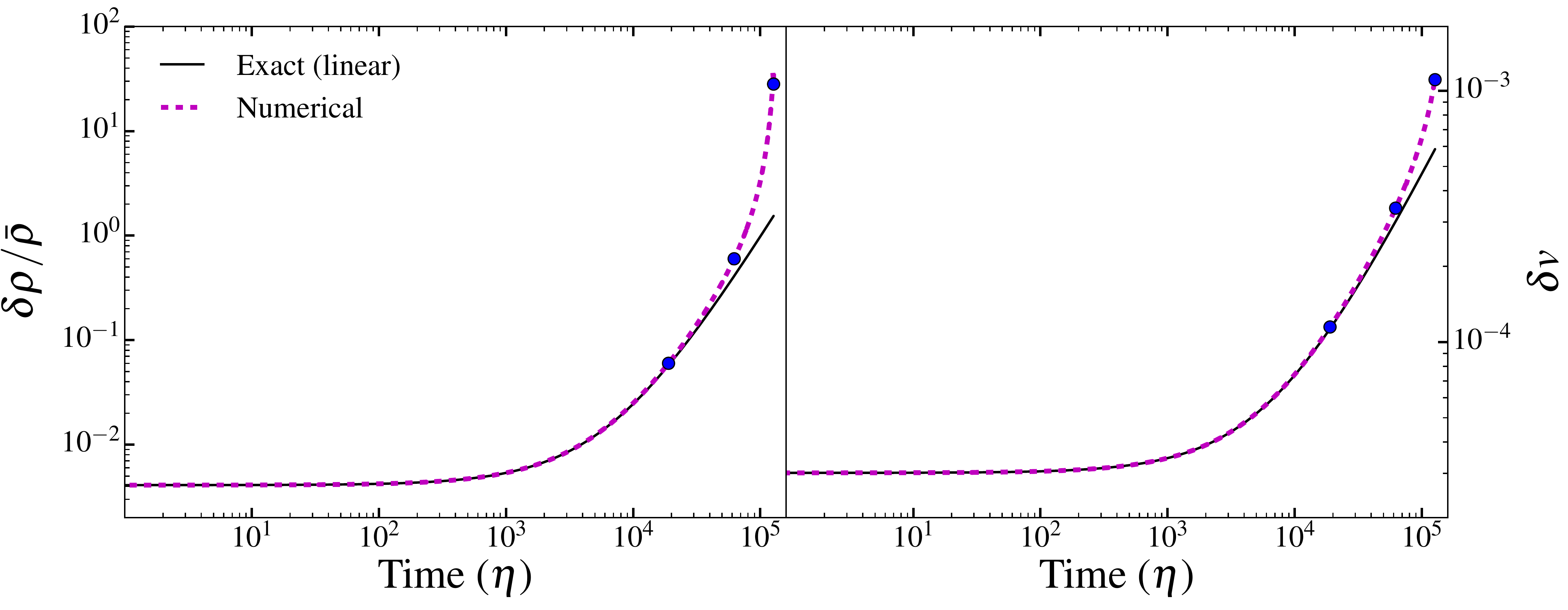}
	\caption{\label{fig:nonlinear_dr_dv}Nonlinear growth of the density (left) and velocity (right) perturbations. Dashed magenta curves show the maximum value within the domain as a function of conformal time $\eta$ (in code units), while black solid curves show the analytic solutions for linear growth. Here we show the simulation with domain size $40^{3}$. Blue circles represent the times of the ($\eta>0$) panels shown in Figure \ref{fig:nonlinear_x}.}
\end{figure*}

\section{\label{sec:nonlinear}Nonlinear evolution}
In order to evolve our perturbations to nonlinear amplitude in a reasonable computational time, we increase the size of our initial perturbations to $\Phi_{0}=10^{-6}$, which in turn gives $\delta\sim 10^{-3}$ and $\delta v^{i}\sim 10^{-5}$. The linear approximation remains valid.

We choose the starting redshift to be that of the cosmic microwave background (CMB). That is, we set $z=1000$, such that our initial density perturbation is roughly consistent with the amplitude of temperature fluctuations in the CMB ($\sim 10^{-5}$) \citep{bennett2013}. We emphasise that this redshift, and all redshifts shown in figures, should not be taken literally; its purpose is to assign an approximate \textit{change} in redshift, calculated directly from the FLRW scale factor.

\subsection{\label{sec:nonlinear_results}Results}
Figure~\ref{fig:nonlinear_x} shows a series of one-dimensional slices through the origin of the $y$ and $z$ axes at four different times. Dashed magenta curves show solutions for the density (top row) and velocity (bottom row) perturbations, which may be compared to the black solid curves showing the analytic solutions for linear perturbations. At $\eta=0$ and $\eta=1.9\times10^{4}$ (first and second columns respectively) the solutions are linear, while at $\eta=6.2\times10^{4}$  (third column) both the density and velocity perturbations deviate from linear theory. The perturbations are nonlinear at $\eta=1.3\times10^{5}$ (fourth column) where matter collapses towards the overdensity, indicated by the shift in the maximum velocity. 

The final column shows an apparent decrease in the average density. This is an artefact of taking a one-dimensional slice through a three-dimensional box. Figure~\ref{fig:nonlinear_2D_delta} shows the column-density perturbation, $\delta_{col}$, computed by integrating the density perturbation along the $z$ axis. Panels show $\eta=0$ and $\eta=1.3\times10^{5}$ respectively. The right panel shows an increase of $\sim 3000$ times in the column-density perturbation at $x,y\approx-120,120$. A corresponding void can be seen in the lower right of Figure~\ref{fig:nonlinear_2D_delta}, explaining the underdensity along the $y$ axis seen in the final column of Figure~\ref{fig:nonlinear_x}.

Figure~\ref{fig:nonlinear_dr_dv} shows the maximum value of the density (left) and velocity (right) perturbations as a function of time. Dashed magenta curves show the numerical solutions, which may be compared to the black curves showing the linear analytic solutions. 
Perturbations can be seen to deviate from the linear approximation at $\eta\approx3\times10^{4}$, when $\delta\rho/\bar{\rho} \approx 0.1$. 
At $\eta\approx10^{5}$, the maximum of the density and velocity perturbations have respectively grown 25 and 2 times larger than the linear solutions. 

\subsection{\label{sec:nonlinear_highorder}Gravitational slip and tensor perturbations}
Gravitational slip is defined as the difference between the two potentials $\Phi$ and $\Psi$ \citep{daniel2008,daniel2009}, which is zero in the linear regime, see equation \eqref{eq:einstein_4}, but nonzero in the nonlinear regime \citep[see e.g.][]{ballesteros2012}.
We reconstruct $\Phi$ and $\Psi$ from the metric components, although we note the interpretation of these potentials becomes unclear in the nonlinear regime. From \eqref{eq:perturbed_metric_svt} the spatial metric is
\begin{equation}
	\gamma_{ij} = a^{2}\left[(1-2\Phi)\delta_{ij} + h_{ij}\right],
\end{equation}
and we adopt the traceless gauge condition $\delta^{ij}h_{ij}=0$ \citep{green2012,adamek2013}. The potential $\Phi$ is then
\begin{equation}
	\Phi=\frac{1}{2}\left(1 - \frac{\delta^{ij}\gamma_{ij}}{3\,a^{2}}\right), \label{eq:phi_reconstruct}
\end{equation}
which holds for all times the metric \eqref{eq:perturbed_metric_svt} applies. The potential $\Psi$ is more complicated: our gauge choice implies lapse evolution according to \eqref{eq:harmonic_gauge}, where we have set $f(\alpha)=1/4$, and 
\begin{equation}
	K=-3\frac{\dot{a}}{a\alpha},
\end{equation}
in the linear regime, which gives
\begin{equation}
	\dot{\alpha} = \frac{3}{4}\frac{\dot{a}\,\alpha}{a}.
\end{equation}
Integrating this results in a lapse evolution of
\begin{equation}
	\frac{\alpha}{\alpha_{\mathrm{init}}} = D(x^{i}) \left(\frac{a}{a_{\mathrm{init}}}\right)^{3/4},
\end{equation}
where $D(x^{i})$ is a function of our spatial coordinates. According to the metric \eqref{eq:perturbed_metric}, and with $\alpha_{\mathrm{init}}=a_{\mathrm{init}}=1$ this implies \begin{equation}
	\alpha = \sqrt{1+2\Psi}\,a^{3/4},
\end{equation}
from which we reconstruct the potential $\Psi$ to be
\begin{equation}
	\Psi = \frac{1}{2}\left[\left(\frac{\alpha}{a^{3/4}}\right)^{2} - 1\right], \label{eq:psi_reconstruct}
\end{equation}
valid in the linear regime. Our gauge choice $\beta^{i}=0$ implies that in the nonlinear regime we expect additional modes to be present in this reconstruction of $\Psi$.
\begin{figure*}[!ht]
	\includegraphics[width=\textwidth]{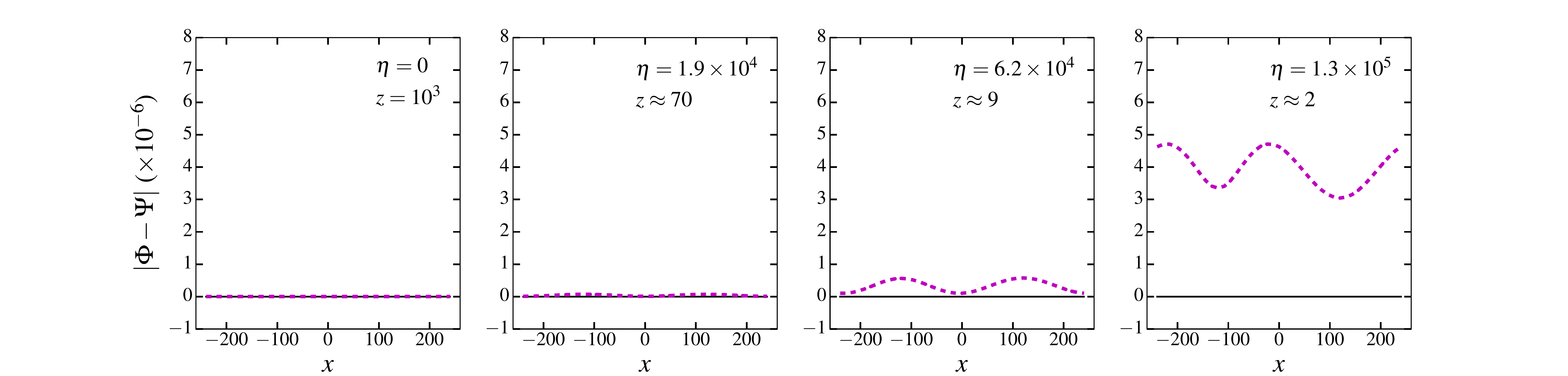} 
	\caption{\label{fig:nonlinear_slip_x}Time evolution of a one-dimensional slice of the gravitational slip. Dashed magenta curves show numerical solutions for a $40^{3}$ domain, while black solid lines show the solution for linear perturbation theory; zero. The potentials $\Phi$ and $\Psi$ are reconstructed according to \eqref{eq:phi_reconstruct} and \eqref{eq:psi_reconstruct} respectively. Initial data ($\eta=0$) is shown in the left column, and time increases towards the right as indicated by timestamps. We show the simulation redshift as an indicator of the approximate \textit{change} in redshift only, and all quantities here are shown in code units.}
\end{figure*}

We use an FLRW simulation for the scale factor $a$ in \eqref{eq:phi_reconstruct} and \eqref{eq:psi_reconstruct}, from which we calculate the gravitational slip $|\Phi - \Psi|$.
This is potentially problematic once the perturbations become nonlinear, as the gauges of the two simulations will differ.
Figure~\ref{fig:nonlinear_slip_x} shows one-dimensional slices of the gravitational slip at the same times as was shown in Figure~\ref{fig:nonlinear_x}. Dashed curves show the numerical results, with black lines showing the linear solution; zero gravitational slip. In the fourth panel ($\eta=1.3\times10^{5}$) we see a positive shift of the gravitational slip to $\approx 4\times10^{-6}$ for this one-dimensional slice, with the maximum value in the three-dimensional domain being $6.5\times10^{-6}$ at this time. The Newtonian potential $\Psi$ has a positive average value at $\eta=1.3\times10^{5}$, due to the majority of the domain being underdense (see Figure~\ref{fig:nonlinear_2D_delta}), and the potential $\Phi$ takes a negative average value. This can be interpreted as an overall positive contribution to the expansion, from the metric \eqref{eq:perturbed_metric_svt}.

Relativistic corrections to one-dimensional N-body simulations in \citep{adamek2013} resulted in a gravitational slip of $4\times10^{-6}$. We show a gravitational slip of the same amplitude, including the full effects of general relativity in a three-dimensional simulation, for a time when our density perturbation is comparable in size to that of \citep{adamek2013}.
Gravitational slip is a measurable effect that can be quantified by combining weak gravitational lensing and galaxy clustering \citep{bertschinger2011}. 
Our simulations show tentative evidence for the importance of gravitational slip due to nonlinear gravitational effects. However, robust predictions require higher resolution simulations with more realistic initial conditions.
\begin{figure}[h]
	\includegraphics[width=\columnwidth]{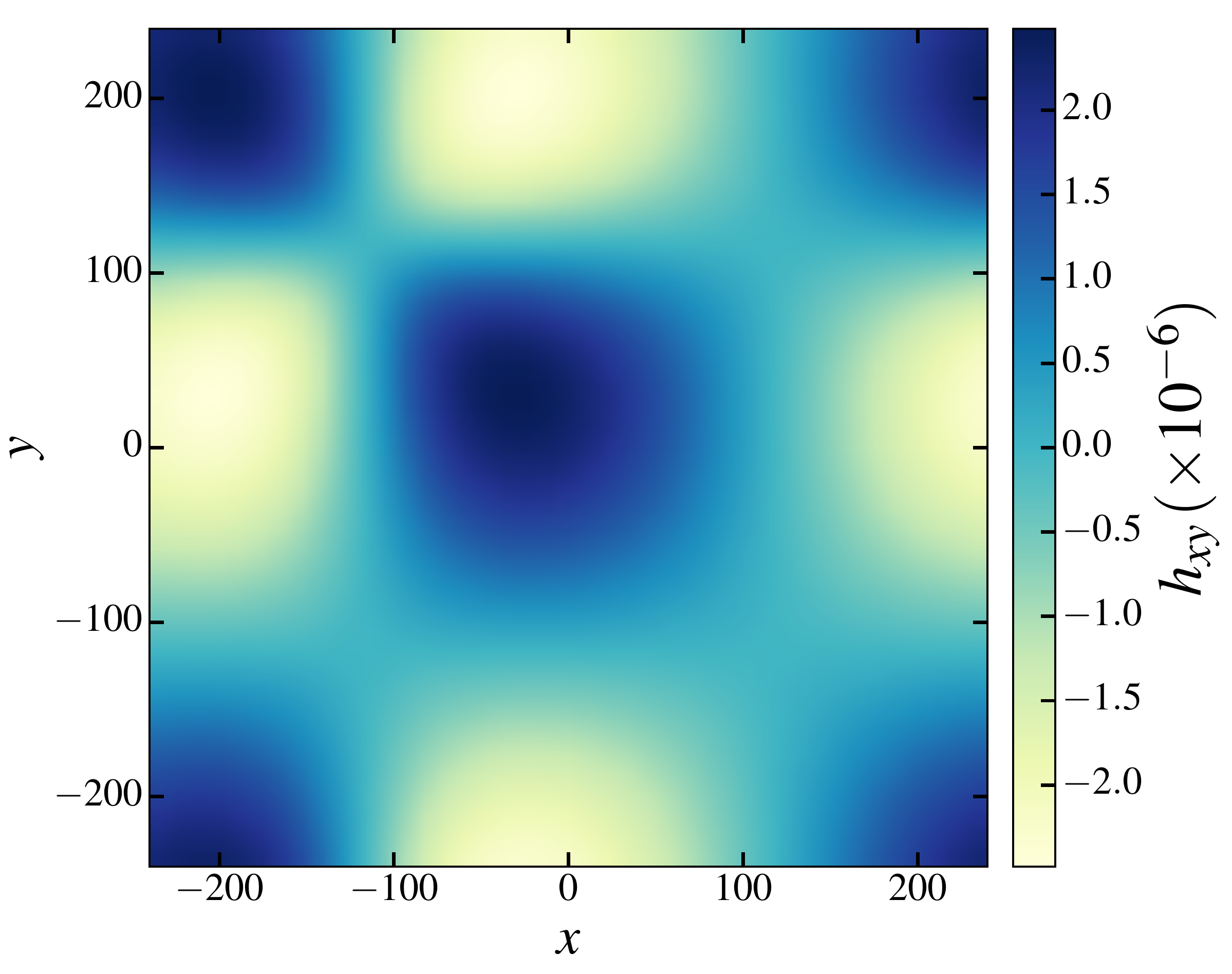}
	\caption{\label{fig:hxy_2D}Two-dimensional slice of the $xy$ component of the tensor perturbation $h_{ij}$ at $\eta=1.3\times10^{5}$. We use \eqref{eq:hij_recon} to calculate $h_{xy}$ using the off-diagonal metric component $g_{xy}$. All quantities are shown in code units for our $40^{3}$ simulation. }
\end{figure}

In our initial conditions we neglected vector and tensor perturbations in the perturbed FLRW metric \eqref{eq:perturbed_metric_svt}, since in the linear regime the scalar perturbations dominate. These higher order perturbations appear in the nonlinear regime. The tensor perturbation can be extracted from the off-diagonal, spatial components of the metric,
\begin{equation}\label{eq:hij_recon}
	\gamma_{ij} = a^{2}h_{ij} \quad \mathrm{for}\,i\neq j,
\end{equation}
however, details of these tensor modes may be dependent on the choice of gauge. 
We calculate $h_{ij}$ using the value of $a$ as per the scalar perturbations.

Figure~\ref{fig:hxy_2D} shows a two-dimensional cross-section of the $xy$ component of the tensor perturbation $h_{ij}$. All other components are identical. The cross-section is shown at $\eta=1.3\times10^{5}$, corresponding to the right panel of Figures~\ref{fig:nonlinear_x}, \ref{fig:nonlinear_2D_delta} and \ref{fig:nonlinear_slip_x}. While the maximum amplitude of the tensor perturbation is small ($\sim 2\times10^{-6}$), 
an asymmetry develops in $h_{xy}$, corresponding to the location of the overdensity in Figure~\ref{fig:nonlinear_2D_delta}.  We also see a diffusion of the tensor perturbation in the void, indicating the beginning of growth of higher order perturbations. 


\section{\label{sec:conclusion}Discussion and Conclusions}
We have demonstrated the feasibility of inhomogeneous cosmological simulations in full general relativity using the Einstein Toolkit. The overall approach is similar to other recent attempts \citep{giblin2016a,bentivegna2015}, with the main difference being in the construction of initial conditions which allows us to simulate a pure growing mode, instead of a mix of growing and decaying modes \citep[see][]{daverio2016}. We also use a different code to \citep{giblin2016a}, allowing for independent verification. As with the other studies we were able to demonstrate the evolution of a density perturbation into the nonlinear regime. 

As this is a preliminary study, we have focused on the numerical accuracy and convergence, rather than a detailed investigation of physical effects such as backreaction. Our main conclusions are:
\begin{enumerate}
\item We demonstrate fourth-order convergence of the numerical solution to the exact solution for a flat, dust FLRW universe with errors $\sim 10^{-5}$ even at low spatial resolution ($40^{3}$).
\item We demonstrate second-order convergence of the numerical solutions for the growth of linear perturbations, matching the analytic solutions for the cosmic evolution of density, velocity and metric perturbations to within $\sim10^{-3}$.
\item We show that numerical relativity can successfully be used to follow the formation of cosmological structures into the nonlinear regime. We demonstrate the appearance of non-zero gravitational slip and tensor modes once perturbations are nonlinear with amplitudes of $\sim4\times10^{-6}$ and $\sim2\times10^{-6}$ respectively. \end{enumerate}

The main limitation to our study is that we have employed only low-resolution simulations compared to current Newtonian N-body cosmological simulations \citep[e.g.][]{genel2014,springel2005,kim2011}, and used only simple initial conditions rather than a more realistic spectrum of perturbations \citep[but see][]{giblin2016a}. Representing the density field on a grid means our simulations are limited by the formation of shell-crossing singularities. The relative computational expense means that general relativistic simulations are unlikely to replace the Newtonian approach in the near future. However, they are an important check on the accuracy of the approximations employed.

\section*{Acknowledgements}
We thank the anonymous referee for comments that helped improve the paper.
We thank Bruno Giacomazzo and the organisers of the Einstein Toolkit Summer Workshop in June 2016 for their advice and support, specifically Eloisa Bentivegna for valuable conversations and Wolfgang Kastaun for support with his visualisation package for Cactus. We thank Todd Oliynyk, Krzysztof Bolejko, Tamara Davis, Chris Blake, Greg Poole, David Wiltshire and Robert Wald for useful discussions. We used the Riemannian Geometry \& Tensor Calculus (RGTC) package for Mathematica, written by Sotirios Bonanos. Our simulations were performed on the Multi-modal Australian ScienceS Imaging and Visualisation Environment (MASSIVE) M2 facility located at Monash University. DJP gratefully acknowledges funding from the Australian Research Council (ARC) via Future Fellowship FT130100034.  PDL gratefully acknowledges funding from the ARC via FT160100112 and Discovery Project DP1410102578.


\appendix
\section{Newtonian Gauge} \label{appx:newt_gauge}
\begin{figure*}[!ht]
	\includegraphics[width=\textwidth]{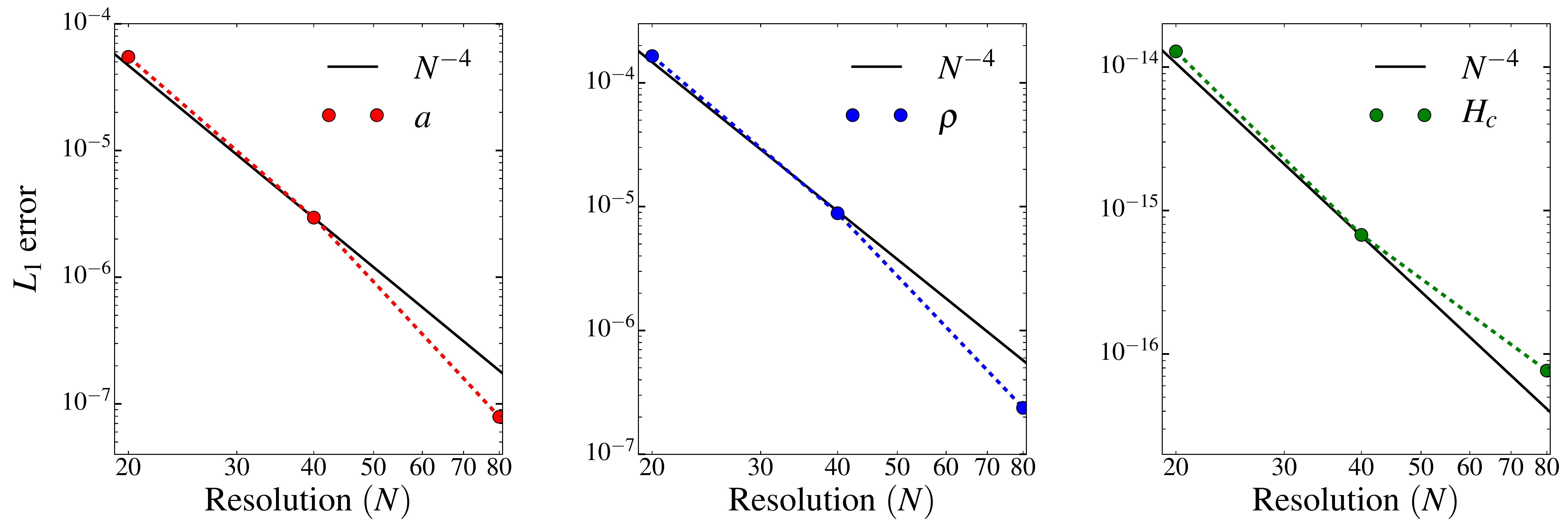}
	\caption{\label{fig:FLRW_RMS_newt}Fourth-order convergence of the FLRW solutions analysed in the Newtonian gauge. We show $L_{1}$ error as a function of resolution for the scale factor (left), density (middle), and Hamiltonian constraint (right). $N$ refers to the number of grid points along one spatial dimension. Filled circles indicate data points from the simulations, and black solid lines indicate the expected $N^{-4}$ convergence.}
\end{figure*}

Throughout this paper we work in the longitudinal gauge. For completeness, we show here the equivalent background and perturbation equations in the Newtonian gauge.
The flat FLRW metric is
\begin{equation}
	ds^{2} =  - d\tau^{2} + a^{2}(\tau)dx^{i}dx^{j}\delta_{ij},
\end{equation}
which gives the Friedmann equations for a dust ($P\ll\rho$) universe to be
\begin{subequations}\label{eq:Friedmann_newt}
	\begin{align}
		\left(\frac{a'}{a}\right)^{2} &= \frac{8\pi\rho}{3}, \label{eq:friedmann_1_newt}\\
		\rho' &= -3\frac{a'}{a}\rho, \label{eq:friedmann_2_newt}
	\end{align}
\end{subequations}
where a dash represents $d/d\tau$.
Solutions to these give the familiar time dependence of the scale factor,
\begin{equation}
		\frac{a}{a_{\mathrm{init}}} = s^{2/3},\quad
		\frac{\rho}{\rho_{\mathrm{init}}} = s^{-2},
\end{equation}
where 
\begin{equation}
	s\equiv1+\sqrt{6\pi\rho^{*}}\tau.
\end{equation}
We match our numerical evolution to this alternative set of solutions by instead making the coordinate transform $t=t(\tau)$. With this we see the expected fourth-order convergence and maximum errors in the scale factor and density of $\sim10^{-7}$ for our highest resolution ($80^{3}$) simulation. Figure~\ref{fig:FLRW_RMS_newt} shows the convergence of the scale factor (left), density (middle) and Hamiltonian constraint (right) for analysis performed in this gauge.

The linearly perturbed FLRW metric in this gauge, including only scalar perturbations, is
\begin{equation}
	ds^{2} = -(1+2\psi)d\tau^{2} + a^{2}(\tau)(1-2\phi)\delta_{ij}dx^{i}dx^{j}, \label{eq:perturbed_metric_newt}
\end{equation}
where $\psi, \phi$ are not the usual gauge-invariant Bardeen potentials (which are defined in the longitudinal gauge).
Solving the perturbed Einstein equations \eqref{eq:perturbed_einstein} in this gauge using the time-time, time-space, trace and trace free components gives
\begin{subequations} \label{eqs:perturbed_einstein_newt}
	\begin{align}
		\nabla^{2}\phi - 3aa'\left(\phi' + \frac{a'}{a} \psi\right) &= 4\pi  \bar{\rho}\,\delta a^{2}, \label{eq:einstein_1_newt} \\ 
		\frac{a'}{a} \partial_{i}\psi + \partial_{i}\phi' &= -4\pi \bar{\rho} \,a^{2} \delta_{ij}\delta v^{j}, \label{eq:einstein_2_newt} \\ 
		\phi'' + \frac{a'}{a}\left(\psi' + 3\phi'\right) &= \frac{1}{2a^{2}}\nabla^{2}(\phi - \psi), \label{eq:einstein_3_newt} \\ 
		\partial_{\langle i}\partial_{j\rangle} \left(\phi - \psi\right) &= 0, \label{eq:einstein_4_newt}
	\end{align}	
\end{subequations}
in the linear regime. Solving these equations we find the form of the potential $\phi$ to be
\begin{equation}
	\phi = f(x^{i}) - \frac{3}{5}s^{-5/3}\,g(x^{i}),
\end{equation}
where $f, g$ are functions of the spatial coordinates. From this we find the density and velocity perturbations to be, respectively,
\begin{subequations}
	\begin{align}
		\delta &= C_{1}\, s^{2/3}\,\nabla^{2}f(x^{i}) - 2\,f(x^{i}) \\
		\phantom{\delta}&\phantom{=C_{1}\,} + 3\,C_{2}\, s^{-1}\,\nabla^{2}g(x^{i}) - \frac{9\,a_{\mathrm{init}}^{3}}{5}s^{-5/3}\,g(x^{i}), \nonumber \\
		\delta v^{i} &= C_{3}\,s^{-1/3}\,\nabla^{i}f(x^{i}) + 3\,C_{4} \,s^{-2}\,\nabla^{i}g(x^{i}), 
	\end{align}
\end{subequations}
where the $C$'s were defined in \eqref{eq:c1c2} and \eqref{eq:c3c4}. 
We set $g(x^{i})=0$ to extract only the growing mode of the density perturbation, giving exact solutions to be
\begin{subequations}
	\begin{align}
		\phi &= f(x^{i}),\\
		\delta &= C_{1}\, s^{2/3}\,\nabla^{2}f(x^{i}) - 2\,f(x^{i}), \label{eq:delta_ex_newt}\\  
		\delta v^{i} &= C_{3}\,s^{-1/3}\,\nabla^{i}f(x^{i}).\label{eq:deltav_ex_newt}
	\end{align}
\end{subequations}
We note that these solutions give equivalent initial conditions to those found in Section \ref{sec:linear_setup} since, initially, $s=\xi=1$. 

\begin{figure*}[!ht]
	\includegraphics[width=0.75\textwidth]{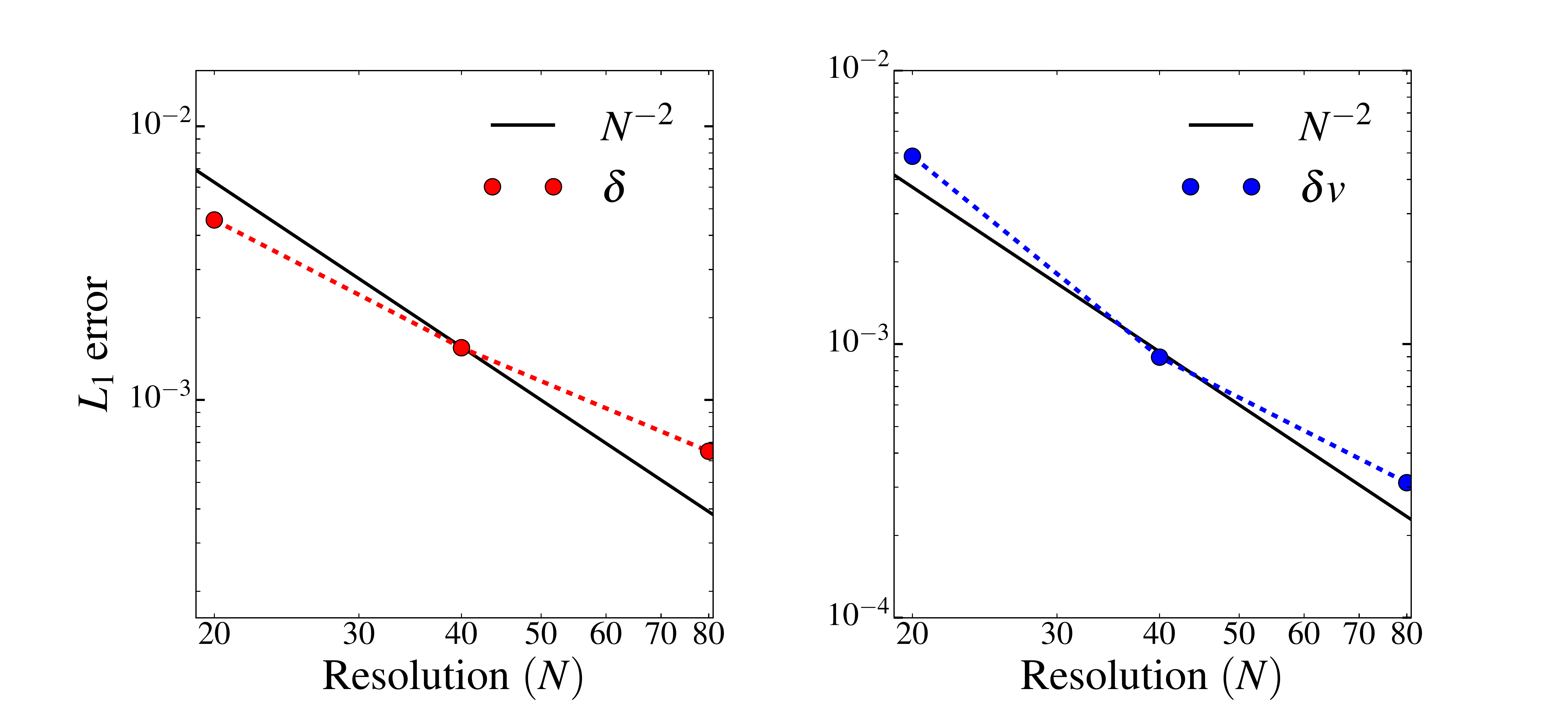}
	\caption{\label{fig:perturb_converge_newt}Second order convergence of the numerical solutions for a linearly perturbed FLRW spacetime, analysed in the Newtonian gauge. We show $L_{1}$ errors in the density (left) and velocity perturbations (right). $N$ refers to the number of grid points along one spatial dimension. Filled circles indicate data points from our simulations, and black solid lines indicate the expected $N^{-2}$ convergence.}
\end{figure*}

We compare our numerical relativity solutions to the exact solutions for linear perturbations in this gauge using the coordinate transform $t=t(\tau)$. We find the expected second-order convergence with maximum errors in the density and velocity perturbations of $\sim10^{-3}$ for our highest resolution ($80^{3}$) simulation. Figure~\ref{fig:perturb_converge_newt} shows the convergence of the density (left) and velocity (right) perturbations when analysed in the Newtonian gauge.


%
%

\bibliography{litreview}

\begin{thebibliography}{77}
\expandafter\ifx\csname natexlab\endcsname\relax\def\natexlab#1{#1}\fi
\expandafter\ifx\csname bibnamefont\endcsname\relax
  \def\bibnamefont#1{#1}\fi
\expandafter\ifx\csname bibfnamefont\endcsname\relax
  \def\bibfnamefont#1{#1}\fi
\expandafter\ifx\csname citenamefont\endcsname\relax
  \def\citenamefont#1{#1}\fi
\expandafter\ifx\csname url\endcsname\relax
  \def\url#1{\texttt{#1}}\fi
\expandafter\ifx\csname urlprefix\endcsname\relax\def\urlprefix{URL }\fi
\providecommand{\bibinfo}[2]{#2}
\providecommand{\eprint}[2][]{\url{#2}}

\bibitem[{\citenamefont{{Genel} et~al.}(2014)\citenamefont{{Genel},
  {Vogelsberger}, {Springel}, {Sijacki}, {Nelson}, {Snyder}, {Rodriguez-Gomez},
  {Torrey}, and {Hernquist}}}]{genel2014}
\bibinfo{author}{\bibfnamefont{S.}~\bibnamefont{{Genel}}},
  \bibinfo{author}{\bibfnamefont{M.}~\bibnamefont{{Vogelsberger}}},
  \bibinfo{author}{\bibfnamefont{V.}~\bibnamefont{{Springel}}},
  \bibinfo{author}{\bibfnamefont{D.}~\bibnamefont{{Sijacki}}},
  \bibinfo{author}{\bibfnamefont{D.}~\bibnamefont{{Nelson}}},
  \bibinfo{author}{\bibfnamefont{G.}~\bibnamefont{{Snyder}}},
  \bibinfo{author}{\bibfnamefont{V.}~\bibnamefont{{Rodriguez-Gomez}}},
  \bibinfo{author}{\bibfnamefont{P.}~\bibnamefont{{Torrey}}}, \bibnamefont{and}
  \bibinfo{author}{\bibfnamefont{L.}~\bibnamefont{{Hernquist}}},
  \bibinfo{journal}{\mnras} \textbf{\bibinfo{volume}{445}},
  \bibinfo{pages}{175} (\bibinfo{year}{2014}).

\bibitem[{\citenamefont{{Springel} et~al.}(2005)\citenamefont{{Springel},
  {White}, {Jenkins}, {Frenk}, {Yoshida}, {Gao}, {Navarro}, {Thacker},
  {Croton}, {Helly} et~al.}}]{springel2005}
\bibinfo{author}{\bibfnamefont{V.}~\bibnamefont{{Springel}}},
  \bibinfo{author}{\bibfnamefont{S.~D.~M.} \bibnamefont{{White}}},
  \bibinfo{author}{\bibfnamefont{A.}~\bibnamefont{{Jenkins}}},
  \bibinfo{author}{\bibfnamefont{C.~S.} \bibnamefont{{Frenk}}},
  \bibinfo{author}{\bibfnamefont{N.}~\bibnamefont{{Yoshida}}},
  \bibinfo{author}{\bibfnamefont{L.}~\bibnamefont{{Gao}}},
  \bibinfo{author}{\bibfnamefont{J.}~\bibnamefont{{Navarro}}},
  \bibinfo{author}{\bibfnamefont{R.}~\bibnamefont{{Thacker}}},
  \bibinfo{author}{\bibfnamefont{D.}~\bibnamefont{{Croton}}},
  \bibinfo{author}{\bibfnamefont{J.}~\bibnamefont{{Helly}}},
  \bibnamefont{et~al.}, \bibinfo{journal}{\nat} \textbf{\bibinfo{volume}{435}},
  \bibinfo{pages}{629} (\bibinfo{year}{2005}).

\bibitem[{\citenamefont{{Kim} et~al.}(2011)\citenamefont{{Kim}, {Park},
  {Rossi}, {Lee}, and {Gott}}}]{kim2011}
\bibinfo{author}{\bibfnamefont{J.}~\bibnamefont{{Kim}}},
  \bibinfo{author}{\bibfnamefont{C.}~\bibnamefont{{Park}}},
  \bibinfo{author}{\bibfnamefont{G.}~\bibnamefont{{Rossi}}},
  \bibinfo{author}{\bibfnamefont{S.~M.} \bibnamefont{{Lee}}}, \bibnamefont{and}
  \bibinfo{author}{\bibfnamefont{J.~R.} \bibnamefont{{Gott}},
  \bibfnamefont{III}}, \bibinfo{journal}{Journal of Korean Astronomical
  Society} \textbf{\bibinfo{volume}{44}}, \bibinfo{pages}{217}
  (\bibinfo{year}{2011}).

\bibitem[{\citenamefont{{Yadav} et~al.}(2010)\citenamefont{{Yadav}, {Bagla},
  and {Khandai}}}]{yadav2010}
\bibinfo{author}{\bibfnamefont{J.~K.} \bibnamefont{{Yadav}}},
  \bibinfo{author}{\bibfnamefont{J.~S.} \bibnamefont{{Bagla}}},
  \bibnamefont{and}
  \bibinfo{author}{\bibfnamefont{N.}~\bibnamefont{{Khandai}}},
  \bibinfo{journal}{\mnras} \textbf{\bibinfo{volume}{405}},
  \bibinfo{pages}{2009} (\bibinfo{year}{2010}).

\bibitem[{\citenamefont{{Scrimgeour} et~al.}(2012)\citenamefont{{Scrimgeour},
  {Davis}, {Blake}, {James}, {Poole}, {Staveley-Smith}, {Brough}, {Colless},
  {Contreras}, {Couch} et~al.}}]{scrimgeour2012}
\bibinfo{author}{\bibfnamefont{M.~I.} \bibnamefont{{Scrimgeour}}},
  \bibinfo{author}{\bibfnamefont{T.}~\bibnamefont{{Davis}}},
  \bibinfo{author}{\bibfnamefont{C.}~\bibnamefont{{Blake}}},
  \bibinfo{author}{\bibfnamefont{J.~B.} \bibnamefont{{James}}},
  \bibinfo{author}{\bibfnamefont{G.~B.} \bibnamefont{{Poole}}},
  \bibinfo{author}{\bibfnamefont{L.}~\bibnamefont{{Staveley-Smith}}},
  \bibinfo{author}{\bibfnamefont{S.}~\bibnamefont{{Brough}}},
  \bibinfo{author}{\bibfnamefont{M.}~\bibnamefont{{Colless}}},
  \bibinfo{author}{\bibfnamefont{C.}~\bibnamefont{{Contreras}}},
  \bibinfo{author}{\bibfnamefont{W.}~\bibnamefont{{Couch}}},
  \bibnamefont{et~al.}, \bibinfo{journal}{\mnras}
  \textbf{\bibinfo{volume}{425}}, \bibinfo{pages}{116} (\bibinfo{year}{2012}).

\bibitem[{\citenamefont{{Amendola} et~al.}(2016)\citenamefont{{Amendola},
  {Appleby}, {Avgoustidis} et~al.}}]{amendola2016}
\bibinfo{author}{\bibfnamefont{L.}~\bibnamefont{{Amendola}}},
  \bibinfo{author}{\bibfnamefont{S.}~\bibnamefont{{Appleby}}},
  \bibinfo{author}{\bibfnamefont{A.}~\bibnamefont{{Avgoustidis}}},
  \bibnamefont{et~al.}, \bibinfo{journal}{ArXiv e-prints}
  (\bibinfo{year}{2016}).

\bibitem[{\citenamefont{{Maartens} et~al.}(2015)\citenamefont{{Maartens},
  {Abdalla}, {Jarvis}, {Santos}, and {SKA Cosmology SWG}}}]{maartens2015}
\bibinfo{author}{\bibfnamefont{R.}~\bibnamefont{{Maartens}}},
  \bibinfo{author}{\bibfnamefont{F.~B.} \bibnamefont{{Abdalla}}},
  \bibinfo{author}{\bibfnamefont{M.}~\bibnamefont{{Jarvis}}},
  \bibinfo{author}{\bibfnamefont{M.~G.} \bibnamefont{{Santos}}},
  \bibnamefont{and} \bibinfo{author}{\bibfnamefont{f.~t.} \bibnamefont{{SKA
  Cosmology SWG}}}, \bibinfo{journal}{ArXiv e-prints}  (\bibinfo{year}{2015}).

\bibitem[{\citenamefont{{Ivezic} et~al.}(2008)\citenamefont{{Ivezic}, {Tyson},
  {Abel} et~al.}}]{ivezic2008}
\bibinfo{author}{\bibfnamefont{Z.}~\bibnamefont{{Ivezic}}},
  \bibinfo{author}{\bibfnamefont{J.~A.} \bibnamefont{{Tyson}}},
  \bibinfo{author}{\bibfnamefont{B.}~\bibnamefont{{Abel}}},
  \bibnamefont{et~al.}, \bibinfo{journal}{ArXiv e-prints}
  (\bibinfo{year}{2008}).

\bibitem[{\citenamefont{{R{\"a}s{\"a}nen}}(2004)}]{rasanen2004}
\bibinfo{author}{\bibfnamefont{S.}~\bibnamefont{{R{\"a}s{\"a}nen}}},
  \bibinfo{journal}{\jcap} \textbf{\bibinfo{volume}{2}}, \bibinfo{pages}{003}
  (\bibinfo{year}{2004}).

\bibitem[{\citenamefont{{Kolb} et~al.}(2005)\citenamefont{{Kolb}, {Matarrese},
  {Notari}, and {Riotto}}}]{kolb2005}
\bibinfo{author}{\bibfnamefont{E.~W.} \bibnamefont{{Kolb}}},
  \bibinfo{author}{\bibfnamefont{S.}~\bibnamefont{{Matarrese}}},
  \bibinfo{author}{\bibfnamefont{A.}~\bibnamefont{{Notari}}}, \bibnamefont{and}
  \bibinfo{author}{\bibfnamefont{A.}~\bibnamefont{{Riotto}}},
  \bibinfo{journal}{\prd} \textbf{\bibinfo{volume}{71}},
  \bibinfo{pages}{023524} (\bibinfo{year}{2005}).

\bibitem[{\citenamefont{{Kolb} et~al.}(2006)\citenamefont{{Kolb}, {Matarrese},
  and {Riotto}}}]{kolb2006}
\bibinfo{author}{\bibfnamefont{E.~W.} \bibnamefont{{Kolb}}},
  \bibinfo{author}{\bibfnamefont{S.}~\bibnamefont{{Matarrese}}},
  \bibnamefont{and} \bibinfo{author}{\bibfnamefont{A.}~\bibnamefont{{Riotto}}},
  \bibinfo{journal}{New Journal of Physics} \textbf{\bibinfo{volume}{8}},
  \bibinfo{pages}{322} (\bibinfo{year}{2006}).

\bibitem[{\citenamefont{{Notari}}(2006)}]{notari2006}
\bibinfo{author}{\bibfnamefont{A.}~\bibnamefont{{Notari}}},
  \bibinfo{journal}{Modern Physics Letters A} \textbf{\bibinfo{volume}{21}},
  \bibinfo{pages}{2997} (\bibinfo{year}{2006}).

\bibitem[{\citenamefont{{R{\"a}s{\"a}nen}}(2006{\natexlab{a}})}]{rasanen2006a}
\bibinfo{author}{\bibfnamefont{S.}~\bibnamefont{{R{\"a}s{\"a}nen}}},
  \bibinfo{journal}{Classical and Quantum Gravity}
  \textbf{\bibinfo{volume}{23}}, \bibinfo{pages}{1823}
  (\bibinfo{year}{2006}{\natexlab{a}}).

\bibitem[{\citenamefont{{R{\"a}s{\"a}nen}}(2006{\natexlab{b}})}]{rasanen2006b}
\bibinfo{author}{\bibfnamefont{S.}~\bibnamefont{{R{\"a}s{\"a}nen}}},
  \bibinfo{journal}{\jcap} \textbf{\bibinfo{volume}{11}}, \bibinfo{pages}{3}
  (\bibinfo{year}{2006}{\natexlab{b}}).

\bibitem[{\citenamefont{{Li} and {Schwarz}}(2007)}]{li2007}
\bibinfo{author}{\bibfnamefont{N.}~\bibnamefont{{Li}}} \bibnamefont{and}
  \bibinfo{author}{\bibfnamefont{D.~J.} \bibnamefont{{Schwarz}}},
  \bibinfo{journal}{\prd} \textbf{\bibinfo{volume}{76}},
  \bibinfo{pages}{083011} (\bibinfo{year}{2007}).

\bibitem[{\citenamefont{{Li} and {Schwarz}}(2008)}]{li2008}
\bibinfo{author}{\bibfnamefont{N.}~\bibnamefont{{Li}}} \bibnamefont{and}
  \bibinfo{author}{\bibfnamefont{D.~J.} \bibnamefont{{Schwarz}}},
  \bibinfo{journal}{\prd} \textbf{\bibinfo{volume}{78}},
  \bibinfo{pages}{083531} (\bibinfo{year}{2008}).

\bibitem[{\citenamefont{{Larena} et~al.}(2009)\citenamefont{{Larena}, {Alimi},
  {Buchert}, {Kunz}, and {Corasaniti}}}]{larena2009}
\bibinfo{author}{\bibfnamefont{J.}~\bibnamefont{{Larena}}},
  \bibinfo{author}{\bibfnamefont{J.-M.} \bibnamefont{{Alimi}}},
  \bibinfo{author}{\bibfnamefont{T.}~\bibnamefont{{Buchert}}},
  \bibinfo{author}{\bibfnamefont{M.}~\bibnamefont{{Kunz}}}, \bibnamefont{and}
  \bibinfo{author}{\bibfnamefont{P.-S.} \bibnamefont{{Corasaniti}}},
  \bibinfo{journal}{\prd} \textbf{\bibinfo{volume}{79}},
  \bibinfo{pages}{083011} (\bibinfo{year}{2009}).

\bibitem[{\citenamefont{{Buchert} et~al.}(2015)\citenamefont{{Buchert},
  {Carfora}, {Ellis}, {Kolb}, {MacCallum}, {Ostrowski}, {R{\"a}s{\"a}nen},
  {Roukema}, {Andersson}, {Coley} et~al.}}]{buchert2015}
\bibinfo{author}{\bibfnamefont{T.}~\bibnamefont{{Buchert}}},
  \bibinfo{author}{\bibfnamefont{M.}~\bibnamefont{{Carfora}}},
  \bibinfo{author}{\bibfnamefont{G.~F.~R.} \bibnamefont{{Ellis}}},
  \bibinfo{author}{\bibfnamefont{E.~W.} \bibnamefont{{Kolb}}},
  \bibinfo{author}{\bibfnamefont{M.~A.~H.} \bibnamefont{{MacCallum}}},
  \bibinfo{author}{\bibfnamefont{J.~J.} \bibnamefont{{Ostrowski}}},
  \bibinfo{author}{\bibfnamefont{S.}~\bibnamefont{{R{\"a}s{\"a}nen}}},
  \bibinfo{author}{\bibfnamefont{B.~F.} \bibnamefont{{Roukema}}},
  \bibinfo{author}{\bibfnamefont{L.}~\bibnamefont{{Andersson}}},
  \bibinfo{author}{\bibfnamefont{A.~A.} \bibnamefont{{Coley}}},
  \bibnamefont{et~al.}, \bibinfo{journal}{Classical and Quantum Gravity}
  \textbf{\bibinfo{volume}{32}}, \bibinfo{pages}{215021}
  (\bibinfo{year}{2015}).

\bibitem[{\citenamefont{{Green} and {Wald}}(2016)}]{green2016}
\bibinfo{author}{\bibfnamefont{S.~R.} \bibnamefont{{Green}}} \bibnamefont{and}
  \bibinfo{author}{\bibfnamefont{R.~M.} \bibnamefont{{Wald}}},
  \bibinfo{journal}{Classical and Quantum Gravity}
  \textbf{\bibinfo{volume}{33}}, \bibinfo{pages}{125027}
  (\bibinfo{year}{2016}).

\bibitem[{\citenamefont{{Bolejko} and {Lasky}}(2008)}]{bolejkolasky2008}
\bibinfo{author}{\bibfnamefont{K.}~\bibnamefont{{Bolejko}}} \bibnamefont{and}
  \bibinfo{author}{\bibfnamefont{P.~D.} \bibnamefont{{Lasky}}},
  \bibinfo{journal}{\mnras} \textbf{\bibinfo{volume}{391}},
  \bibinfo{pages}{L59} (\bibinfo{year}{2008}).

\bibitem[{\citenamefont{{Buchert}}(2008)}]{buchert2008}
\bibinfo{author}{\bibfnamefont{T.}~\bibnamefont{{Buchert}}},
  \bibinfo{journal}{General Relativity and Gravitation}
  \textbf{\bibinfo{volume}{40}}, \bibinfo{pages}{467} (\bibinfo{year}{2008}).

\bibitem[{\citenamefont{{Buchert} and {R{\"a}s{\"a}nen}}(2012)}]{buchert2012}
\bibinfo{author}{\bibfnamefont{T.}~\bibnamefont{{Buchert}}} \bibnamefont{and}
  \bibinfo{author}{\bibfnamefont{S.}~\bibnamefont{{R{\"a}s{\"a}nen}}},
  \bibinfo{journal}{Annual Review of Nuclear and Particle Science}
  \textbf{\bibinfo{volume}{62}}, \bibinfo{pages}{57} (\bibinfo{year}{2012}).

\bibitem[{\citenamefont{{Riess} et~al.}(1998)\citenamefont{{Riess},
  {Filippenko}, {Challis}, {Clocchiatti}, {Diercks}, {Garnavich}, {Gilliland},
  {Hogan}, {Jha}, {Kirshner} et~al.}}]{riess1998}
\bibinfo{author}{\bibfnamefont{A.~G.} \bibnamefont{{Riess}}},
  \bibinfo{author}{\bibfnamefont{A.~V.} \bibnamefont{{Filippenko}}},
  \bibinfo{author}{\bibfnamefont{P.}~\bibnamefont{{Challis}}},
  \bibinfo{author}{\bibfnamefont{A.}~\bibnamefont{{Clocchiatti}}},
  \bibinfo{author}{\bibfnamefont{A.}~\bibnamefont{{Diercks}}},
  \bibinfo{author}{\bibfnamefont{P.~M.} \bibnamefont{{Garnavich}}},
  \bibinfo{author}{\bibfnamefont{R.~L.} \bibnamefont{{Gilliland}}},
  \bibinfo{author}{\bibfnamefont{C.~J.} \bibnamefont{{Hogan}}},
  \bibinfo{author}{\bibfnamefont{S.}~\bibnamefont{{Jha}}},
  \bibinfo{author}{\bibfnamefont{R.~P.} \bibnamefont{{Kirshner}}},
  \bibnamefont{et~al.}, \bibinfo{journal}{\aj} \textbf{\bibinfo{volume}{116}},
  \bibinfo{pages}{1009} (\bibinfo{year}{1998}).

\bibitem[{\citenamefont{{Perlmutter} et~al.}(1999)\citenamefont{{Perlmutter},
  {Aldering}, {Goldhaber}, {Knop}, {Nugent}, {Castro}, {Deustua}, {Fabbro},
  {Goobar}, {Groom} et~al.}}]{perlmutter1999}
\bibinfo{author}{\bibfnamefont{S.}~\bibnamefont{{Perlmutter}}},
  \bibinfo{author}{\bibfnamefont{G.}~\bibnamefont{{Aldering}}},
  \bibinfo{author}{\bibfnamefont{G.}~\bibnamefont{{Goldhaber}}},
  \bibinfo{author}{\bibfnamefont{R.~A.} \bibnamefont{{Knop}}},
  \bibinfo{author}{\bibfnamefont{P.}~\bibnamefont{{Nugent}}},
  \bibinfo{author}{\bibfnamefont{P.~G.} \bibnamefont{{Castro}}},
  \bibinfo{author}{\bibfnamefont{S.}~\bibnamefont{{Deustua}}},
  \bibinfo{author}{\bibfnamefont{S.}~\bibnamefont{{Fabbro}}},
  \bibinfo{author}{\bibfnamefont{A.}~\bibnamefont{{Goobar}}},
  \bibinfo{author}{\bibfnamefont{D.~E.} \bibnamefont{{Groom}}},
  \bibnamefont{et~al.}, \bibinfo{journal}{\apj} \textbf{\bibinfo{volume}{517}},
  \bibinfo{pages}{565} (\bibinfo{year}{1999}).

\bibitem[{\citenamefont{{Parkinson} et~al.}(2012)\citenamefont{{Parkinson},
  {Riemer-S{\o}rensen}, {Blake}, {Poole}, {Davis}, {Brough}, {Colless},
  {Contreras}, {Couch}, {Croom} et~al.}}]{parkinson2012}
\bibinfo{author}{\bibfnamefont{D.}~\bibnamefont{{Parkinson}}},
  \bibinfo{author}{\bibfnamefont{S.}~\bibnamefont{{Riemer-S{\o}rensen}}},
  \bibinfo{author}{\bibfnamefont{C.}~\bibnamefont{{Blake}}},
  \bibinfo{author}{\bibfnamefont{G.~B.} \bibnamefont{{Poole}}},
  \bibinfo{author}{\bibfnamefont{T.~M.} \bibnamefont{{Davis}}},
  \bibinfo{author}{\bibfnamefont{S.}~\bibnamefont{{Brough}}},
  \bibinfo{author}{\bibfnamefont{M.}~\bibnamefont{{Colless}}},
  \bibinfo{author}{\bibfnamefont{C.}~\bibnamefont{{Contreras}}},
  \bibinfo{author}{\bibfnamefont{W.}~\bibnamefont{{Couch}}},
  \bibinfo{author}{\bibfnamefont{S.}~\bibnamefont{{Croom}}},
  \bibnamefont{et~al.}, \bibinfo{journal}{\prd} \textbf{\bibinfo{volume}{86}},
  \bibinfo{pages}{103518} (\bibinfo{year}{2012}).

\bibitem[{\citenamefont{{Samushia} et~al.}(2013)\citenamefont{{Samushia},
  {Reid}, {White}, {Percival}, {Cuesta}, {Lombriser}, {Manera}, {Nichol},
  {Schneider}, {Bizyaev} et~al.}}]{samushia2013}
\bibinfo{author}{\bibfnamefont{L.}~\bibnamefont{{Samushia}}},
  \bibinfo{author}{\bibfnamefont{B.~A.} \bibnamefont{{Reid}}},
  \bibinfo{author}{\bibfnamefont{M.}~\bibnamefont{{White}}},
  \bibinfo{author}{\bibfnamefont{W.~J.} \bibnamefont{{Percival}}},
  \bibinfo{author}{\bibfnamefont{A.~J.} \bibnamefont{{Cuesta}}},
  \bibinfo{author}{\bibfnamefont{L.}~\bibnamefont{{Lombriser}}},
  \bibinfo{author}{\bibfnamefont{M.}~\bibnamefont{{Manera}}},
  \bibinfo{author}{\bibfnamefont{R.~C.} \bibnamefont{{Nichol}}},
  \bibinfo{author}{\bibfnamefont{D.~P.} \bibnamefont{{Schneider}}},
  \bibinfo{author}{\bibfnamefont{D.}~\bibnamefont{{Bizyaev}}},
  \bibnamefont{et~al.}, \bibinfo{journal}{\mnras}
  \textbf{\bibinfo{volume}{429}}, \bibinfo{pages}{1514} (\bibinfo{year}{2013}).

\bibitem[{\citenamefont{{Matarrese} and {Terranova}}(1996)}]{matarrese1996}
\bibinfo{author}{\bibfnamefont{S.}~\bibnamefont{{Matarrese}}} \bibnamefont{and}
  \bibinfo{author}{\bibfnamefont{D.}~\bibnamefont{{Terranova}}},
  \bibinfo{journal}{\mnras} \textbf{\bibinfo{volume}{283}},
  \bibinfo{pages}{400} (\bibinfo{year}{1996}).

\bibitem[{\citenamefont{{R{\"a}s{\"a}nen}}(2010)}]{rasanen2010}
\bibinfo{author}{\bibfnamefont{S.}~\bibnamefont{{R{\"a}s{\"a}nen}}},
  \bibinfo{journal}{\prd} \textbf{\bibinfo{volume}{81}},
  \bibinfo{pages}{103512} (\bibinfo{year}{2010}).

\bibitem[{\citenamefont{{Green} and {Wald}}(2011)}]{green2011}
\bibinfo{author}{\bibfnamefont{S.~R.} \bibnamefont{{Green}}} \bibnamefont{and}
  \bibinfo{author}{\bibfnamefont{R.~M.} \bibnamefont{{Wald}}},
  \bibinfo{journal}{\prd} \textbf{\bibinfo{volume}{83}},
  \bibinfo{pages}{084020} (\bibinfo{year}{2011}).

\bibitem[{\citenamefont{{Green} and {Wald}}(2012)}]{green2012}
\bibinfo{author}{\bibfnamefont{S.~R.} \bibnamefont{{Green}}} \bibnamefont{and}
  \bibinfo{author}{\bibfnamefont{R.~M.} \bibnamefont{{Wald}}},
  \bibinfo{journal}{\prd} \textbf{\bibinfo{volume}{85}},
  \bibinfo{pages}{063512} (\bibinfo{year}{2012}).

\bibitem[{\citenamefont{{Adamek} et~al.}(2013)\citenamefont{{Adamek},
  {Daverio}, {Durrer}, and {Kunz}}}]{adamek2013}
\bibinfo{author}{\bibfnamefont{J.}~\bibnamefont{{Adamek}}},
  \bibinfo{author}{\bibfnamefont{D.}~\bibnamefont{{Daverio}}},
  \bibinfo{author}{\bibfnamefont{R.}~\bibnamefont{{Durrer}}}, \bibnamefont{and}
  \bibinfo{author}{\bibfnamefont{M.}~\bibnamefont{{Kunz}}},
  \bibinfo{journal}{\prd} \textbf{\bibinfo{volume}{88}},
  \bibinfo{pages}{103527} (\bibinfo{year}{2013}).

\bibitem[{\citenamefont{{Adamek}
  et~al.}(2016{\natexlab{a}})\citenamefont{{Adamek}, {Daverio}, {Durrer}, and
  {Kunz}}}]{adamek2016a}
\bibinfo{author}{\bibfnamefont{J.}~\bibnamefont{{Adamek}}},
  \bibinfo{author}{\bibfnamefont{D.}~\bibnamefont{{Daverio}}},
  \bibinfo{author}{\bibfnamefont{R.}~\bibnamefont{{Durrer}}}, \bibnamefont{and}
  \bibinfo{author}{\bibfnamefont{M.}~\bibnamefont{{Kunz}}},
  \bibinfo{journal}{\jcap} \textbf{\bibinfo{volume}{7}}, \bibinfo{pages}{053}
  (\bibinfo{year}{2016}{\natexlab{a}}).

\bibitem[{\citenamefont{{Adamek}
  et~al.}(2016{\natexlab{b}})\citenamefont{{Adamek}, {Daverio}, {Durrer}, and
  {Kunz}}}]{adamek2016b}
\bibinfo{author}{\bibfnamefont{J.}~\bibnamefont{{Adamek}}},
  \bibinfo{author}{\bibfnamefont{D.}~\bibnamefont{{Daverio}}},
  \bibinfo{author}{\bibfnamefont{R.}~\bibnamefont{{Durrer}}}, \bibnamefont{and}
  \bibinfo{author}{\bibfnamefont{M.}~\bibnamefont{{Kunz}}},
  \bibinfo{journal}{Nature Physics} \textbf{\bibinfo{volume}{12}},
  \bibinfo{pages}{346} (\bibinfo{year}{2016}{\natexlab{b}}).

\bibitem[{\citenamefont{{Sanghai} and {Clifton}}(2015)}]{sanghai2015}
\bibinfo{author}{\bibfnamefont{V.~A.~A.} \bibnamefont{{Sanghai}}}
  \bibnamefont{and}
  \bibinfo{author}{\bibfnamefont{T.}~\bibnamefont{{Clifton}}},
  \bibinfo{journal}{\prd} \textbf{\bibinfo{volume}{91}},
  \bibinfo{pages}{103532} (\bibinfo{year}{2015}).

\bibitem[{\citenamefont{{Oliynyk}}(2014)}]{oliynyk2014}
\bibinfo{author}{\bibfnamefont{T.~A.} \bibnamefont{{Oliynyk}}},
  \bibinfo{journal}{\prd} \textbf{\bibinfo{volume}{89}},
  \bibinfo{pages}{124002} (\bibinfo{year}{2014}).

\bibitem[{\citenamefont{{Noh} and {Hwang}}(2004)}]{noh2004}
\bibinfo{author}{\bibfnamefont{H.}~\bibnamefont{{Noh}}} \bibnamefont{and}
  \bibinfo{author}{\bibfnamefont{J.-C.} \bibnamefont{{Hwang}}},
  \bibinfo{journal}{\prd} \textbf{\bibinfo{volume}{69}},
  \bibinfo{pages}{104011} (\bibinfo{year}{2004}).

\bibitem[{\citenamefont{{Pretorius}}(2005)}]{pretorius2005}
\bibinfo{author}{\bibfnamefont{F.}~\bibnamefont{{Pretorius}}},
  \bibinfo{journal}{Physical Review Letters} \textbf{\bibinfo{volume}{95}},
  \bibinfo{eid}{121101} (\bibinfo{year}{2005}).

\bibitem[{\citenamefont{{Campanelli} et~al.}(2006)\citenamefont{{Campanelli},
  {Lousto}, {Marronetti}, and {Zlochower}}}]{campanelli2006}
\bibinfo{author}{\bibfnamefont{M.}~\bibnamefont{{Campanelli}}},
  \bibinfo{author}{\bibfnamefont{C.~O.} \bibnamefont{{Lousto}}},
  \bibinfo{author}{\bibfnamefont{P.}~\bibnamefont{{Marronetti}}},
  \bibnamefont{and}
  \bibinfo{author}{\bibfnamefont{Y.}~\bibnamefont{{Zlochower}}},
  \bibinfo{journal}{Physical Review Letters} \textbf{\bibinfo{volume}{96}},
  \bibinfo{pages}{111101} (\bibinfo{year}{2006}).

\bibitem[{\citenamefont{{Baker} et~al.}(2006)\citenamefont{{Baker},
  {Centrella}, {Choi}, {Koppitz}, and {van Meter}}}]{baker2006}
\bibinfo{author}{\bibfnamefont{J.~G.} \bibnamefont{{Baker}}},
  \bibinfo{author}{\bibfnamefont{J.}~\bibnamefont{{Centrella}}},
  \bibinfo{author}{\bibfnamefont{D.-I.} \bibnamefont{{Choi}}},
  \bibinfo{author}{\bibfnamefont{M.}~\bibnamefont{{Koppitz}}},
  \bibnamefont{and} \bibinfo{author}{\bibfnamefont{J.}~\bibnamefont{{van
  Meter}}}, \bibinfo{journal}{Physical Review Letters}
  \textbf{\bibinfo{volume}{96}}, \bibinfo{pages}{111102}
  (\bibinfo{year}{2006}).

\bibitem[{\citenamefont{{Arnowitt} et~al.}(1959)\citenamefont{{Arnowitt},
  {Deser}, and {Misner}}}]{arnowitt1959}
\bibinfo{author}{\bibfnamefont{R.}~\bibnamefont{{Arnowitt}}},
  \bibinfo{author}{\bibfnamefont{S.}~\bibnamefont{{Deser}}}, \bibnamefont{and}
  \bibinfo{author}{\bibfnamefont{C.~W.} \bibnamefont{{Misner}}},
  \bibinfo{journal}{Physical Review} \textbf{\bibinfo{volume}{116}},
  \bibinfo{pages}{1322} (\bibinfo{year}{1959}).

\bibitem[{\citenamefont{{Centrella} and {Matzner}}(1979)}]{centrella1979}
\bibinfo{author}{\bibfnamefont{J.}~\bibnamefont{{Centrella}}} \bibnamefont{and}
  \bibinfo{author}{\bibfnamefont{R.~A.} \bibnamefont{{Matzner}}},
  \bibinfo{journal}{\apj} \textbf{\bibinfo{volume}{230}}, \bibinfo{pages}{311}
  (\bibinfo{year}{1979}).

\bibitem[{\citenamefont{{Centrella}}(1980)}]{centrella1980}
\bibinfo{author}{\bibfnamefont{J.}~\bibnamefont{{Centrella}}},
  \bibinfo{journal}{\prd} \textbf{\bibinfo{volume}{21}}, \bibinfo{pages}{2776}
  (\bibinfo{year}{1980}).

\bibitem[{\citenamefont{{Centrella} and {Matzner}}(1982)}]{centrella1982}
\bibinfo{author}{\bibfnamefont{J.}~\bibnamefont{{Centrella}}} \bibnamefont{and}
  \bibinfo{author}{\bibfnamefont{R.~A.} \bibnamefont{{Matzner}}},
  \bibinfo{journal}{\prd} \textbf{\bibinfo{volume}{25}}, \bibinfo{pages}{930}
  (\bibinfo{year}{1982}).

\bibitem[{\citenamefont{{Centrella} and {Wilson}}(1983)}]{centrella1983}
\bibinfo{author}{\bibfnamefont{J.}~\bibnamefont{{Centrella}}} \bibnamefont{and}
  \bibinfo{author}{\bibfnamefont{J.~R.} \bibnamefont{{Wilson}}},
  \bibinfo{journal}{\apj} \textbf{\bibinfo{volume}{273}}, \bibinfo{pages}{428}
  (\bibinfo{year}{1983}).

\bibitem[{\citenamefont{{Centrella} and {Wilson}}(1984)}]{centrella1984}
\bibinfo{author}{\bibfnamefont{J.}~\bibnamefont{{Centrella}}} \bibnamefont{and}
  \bibinfo{author}{\bibfnamefont{J.~R.} \bibnamefont{{Wilson}}},
  \bibinfo{journal}{\apj} \textbf{\bibinfo{volume}{54}}, \bibinfo{pages}{229}
  (\bibinfo{year}{1984}).

\bibitem[{\citenamefont{{Rekier} et~al.}(2015)\citenamefont{{Rekier},
  {Cordero-Carri{\'o}n}, and {F{\"u}zfa}}}]{rekier2015}
\bibinfo{author}{\bibfnamefont{J.}~\bibnamefont{{Rekier}}},
  \bibinfo{author}{\bibfnamefont{I.}~\bibnamefont{{Cordero-Carri{\'o}n}}},
  \bibnamefont{and}
  \bibinfo{author}{\bibfnamefont{A.}~\bibnamefont{{F{\"u}zfa}}},
  \bibinfo{journal}{\prd} \textbf{\bibinfo{volume}{91}},
  \bibinfo{pages}{024025} (\bibinfo{year}{2015}).

\bibitem[{\citenamefont{{Torres} et~al.}(2014)\citenamefont{{Torres},
  {Alcubierre}, {Diez-Tejedor}, and {N{\'u}{\~n}ez}}}]{torres2014}
\bibinfo{author}{\bibfnamefont{J.~M.} \bibnamefont{{Torres}}},
  \bibinfo{author}{\bibfnamefont{M.}~\bibnamefont{{Alcubierre}}},
  \bibinfo{author}{\bibfnamefont{A.}~\bibnamefont{{Diez-Tejedor}}},
  \bibnamefont{and}
  \bibinfo{author}{\bibfnamefont{D.}~\bibnamefont{{N{\'u}{\~n}ez}}},
  \bibinfo{journal}{\prd} \textbf{\bibinfo{volume}{90}},
  \bibinfo{pages}{123002} (\bibinfo{year}{2014}).

\bibitem[{\citenamefont{{Giblin}
  et~al.}(2016{\natexlab{a}})\citenamefont{{Giblin}, {Mertens}, and
  {Starkman}}}]{giblin2016a}
\bibinfo{author}{\bibfnamefont{J.~T.} \bibnamefont{{Giblin}}},
  \bibinfo{author}{\bibfnamefont{J.~B.} \bibnamefont{{Mertens}}},
  \bibnamefont{and} \bibinfo{author}{\bibfnamefont{G.~D.}
  \bibnamefont{{Starkman}}}, \bibinfo{journal}{Physical Review Letters}
  \textbf{\bibinfo{volume}{116}}, \bibinfo{pages}{251301}
  (\bibinfo{year}{2016}{\natexlab{a}}).

\bibitem[{\citenamefont{{Giblin}
  et~al.}(2016{\natexlab{b}})\citenamefont{{Giblin}, {Mertens}, and
  {Starkman}}}]{giblin2016b}
\bibinfo{author}{\bibfnamefont{J.~T.} \bibnamefont{{Giblin}},
  \bibfnamefont{Jr}}, \bibinfo{author}{\bibfnamefont{J.~B.}
  \bibnamefont{{Mertens}}}, \bibnamefont{and}
  \bibinfo{author}{\bibfnamefont{G.~D.} \bibnamefont{{Starkman}}},
  \bibinfo{journal}{ArXiv e-prints}  (\bibinfo{year}{2016}{\natexlab{b}}).

\bibitem[{\citenamefont{{Bentivegna} and {Bruni}}(2015)}]{bentivegna2015}
\bibinfo{author}{\bibfnamefont{E.}~\bibnamefont{{Bentivegna}}}
  \bibnamefont{and} \bibinfo{author}{\bibfnamefont{M.}~\bibnamefont{{Bruni}}},
  \bibinfo{journal}{ArXiv e-prints}  (\bibinfo{year}{2015}).

\bibitem[{\citenamefont{{Buchert}}(2000)}]{buchert2000a}
\bibinfo{author}{\bibfnamefont{T.}~\bibnamefont{{Buchert}}},
  \bibinfo{journal}{General Relativity and Gravitation}
  \textbf{\bibinfo{volume}{32}}, \bibinfo{pages}{105} (\bibinfo{year}{2000}).

\bibitem[{\citenamefont{{L{\"o}ffler} et~al.}(2012)\citenamefont{{L{\"o}ffler},
  {Faber}, {Bentivegna}, {Bode}, {Diener}, {Haas}, {Hinder}, {Mundim}, {Ott},
  {Schnetter} et~al.}}]{loffler2012}
\bibinfo{author}{\bibfnamefont{F.}~\bibnamefont{{L{\"o}ffler}}},
  \bibinfo{author}{\bibfnamefont{J.}~\bibnamefont{{Faber}}},
  \bibinfo{author}{\bibfnamefont{E.}~\bibnamefont{{Bentivegna}}},
  \bibinfo{author}{\bibfnamefont{T.}~\bibnamefont{{Bode}}},
  \bibinfo{author}{\bibfnamefont{P.}~\bibnamefont{{Diener}}},
  \bibinfo{author}{\bibfnamefont{R.}~\bibnamefont{{Haas}}},
  \bibinfo{author}{\bibfnamefont{I.}~\bibnamefont{{Hinder}}},
  \bibinfo{author}{\bibfnamefont{B.~C.} \bibnamefont{{Mundim}}},
  \bibinfo{author}{\bibfnamefont{C.~D.} \bibnamefont{{Ott}}},
  \bibinfo{author}{\bibfnamefont{E.}~\bibnamefont{{Schnetter}}},
  \bibnamefont{et~al.}, \bibinfo{journal}{Classical and Quantum Gravity}
  \textbf{\bibinfo{volume}{29}}, \bibinfo{pages}{115001}
  (\bibinfo{year}{2012}).

\bibitem[{\citenamefont{{Zilh{\~a}o} and {L{\"o}ffler}}(2013)}]{zilhao2013}
\bibinfo{author}{\bibfnamefont{M.}~\bibnamefont{{Zilh{\~a}o}}}
  \bibnamefont{and}
  \bibinfo{author}{\bibfnamefont{F.}~\bibnamefont{{L{\"o}ffler}}},
  \bibinfo{journal}{International Journal of Modern Physics A}
  \textbf{\bibinfo{volume}{28}}, \bibinfo{pages}{1340014}
  (\bibinfo{year}{2013}).

\bibitem[{\citenamefont{{Daniel} et~al.}(2008)\citenamefont{{Daniel},
  {Caldwell}, {Cooray}, and {Melchiorri}}}]{daniel2008}
\bibinfo{author}{\bibfnamefont{S.~F.} \bibnamefont{{Daniel}}},
  \bibinfo{author}{\bibfnamefont{R.~R.} \bibnamefont{{Caldwell}}},
  \bibinfo{author}{\bibfnamefont{A.}~\bibnamefont{{Cooray}}}, \bibnamefont{and}
  \bibinfo{author}{\bibfnamefont{A.}~\bibnamefont{{Melchiorri}}},
  \bibinfo{journal}{\prd} \textbf{\bibinfo{volume}{77}},
  \bibinfo{pages}{103513} (\bibinfo{year}{2008}).

\bibitem[{\citenamefont{{Daniel} et~al.}(2009)\citenamefont{{Daniel},
  {Caldwell}, {Cooray}, {Serra}, and {Melchiorri}}}]{daniel2009}
\bibinfo{author}{\bibfnamefont{S.~F.} \bibnamefont{{Daniel}}},
  \bibinfo{author}{\bibfnamefont{R.~R.} \bibnamefont{{Caldwell}}},
  \bibinfo{author}{\bibfnamefont{A.}~\bibnamefont{{Cooray}}},
  \bibinfo{author}{\bibfnamefont{P.}~\bibnamefont{{Serra}}}, \bibnamefont{and}
  \bibinfo{author}{\bibfnamefont{A.}~\bibnamefont{{Melchiorri}}},
  \bibinfo{journal}{\prd} \textbf{\bibinfo{volume}{80}},
  \bibinfo{pages}{023532} (\bibinfo{year}{2009}).

\bibitem[{\citenamefont{{Goodale} et~al.}(2003)\citenamefont{{Goodale},
  {Allen}, {Lanfermann}, {Mass{\'o}}, {Radke}, {Seidel}, and {Shalf}}}]{cactus}
\bibinfo{author}{\bibfnamefont{T.}~\bibnamefont{{Goodale}}},
  \bibinfo{author}{\bibfnamefont{G.}~\bibnamefont{{Allen}}},
  \bibinfo{author}{\bibfnamefont{G.}~\bibnamefont{{Lanfermann}}},
  \bibinfo{author}{\bibfnamefont{J.}~\bibnamefont{{Mass{\'o}}}},
  \bibinfo{author}{\bibfnamefont{T.}~\bibnamefont{{Radke}}},
  \bibinfo{author}{\bibfnamefont{E.}~\bibnamefont{{Seidel}}}, \bibnamefont{and}
  \bibinfo{author}{\bibfnamefont{J.}~\bibnamefont{{Shalf}}}, in
  \emph{\bibinfo{booktitle}{{Vector and Parallel Processing -- VECPAR'2002, 5th
  International Conference, Lecture Notes in Computer Science}}}
  (\bibinfo{publisher}{{Springer}}, \bibinfo{address}{{Berlin}},
  \bibinfo{year}{2003}).

\bibitem[{\citenamefont{{Kastaun} and {Galeazzi}}(2015)}]{kastaun2015}
\bibinfo{author}{\bibfnamefont{W.}~\bibnamefont{{Kastaun}}} \bibnamefont{and}
  \bibinfo{author}{\bibfnamefont{F.}~\bibnamefont{{Galeazzi}}},
  \bibinfo{journal}{\prd} \textbf{\bibinfo{volume}{91}},
  \bibinfo{pages}{064027} (\bibinfo{year}{2015}).

\bibitem[{\citenamefont{{Radice} et~al.}(2015)\citenamefont{{Radice},
  {Rezzolla}, and {Galeazzi}}}]{radice2015}
\bibinfo{author}{\bibfnamefont{D.}~\bibnamefont{{Radice}}},
  \bibinfo{author}{\bibfnamefont{L.}~\bibnamefont{{Rezzolla}}},
  \bibnamefont{and}
  \bibinfo{author}{\bibfnamefont{F.}~\bibnamefont{{Galeazzi}}},
  \bibinfo{journal}{ArXiv e-prints}  (\bibinfo{year}{2015}).

\bibitem[{\citenamefont{{Baiotti} et~al.}(2005)\citenamefont{{Baiotti},
  {Hawke}, {Montero}, {L{\"o}ffler}, {Rezzolla}, {Stergioulas}, {Font}, and
  {Seidel}}}]{baiotti2005}
\bibinfo{author}{\bibfnamefont{L.}~\bibnamefont{{Baiotti}}},
  \bibinfo{author}{\bibfnamefont{I.}~\bibnamefont{{Hawke}}},
  \bibinfo{author}{\bibfnamefont{P.~J.} \bibnamefont{{Montero}}},
  \bibinfo{author}{\bibfnamefont{F.}~\bibnamefont{{L{\"o}ffler}}},
  \bibinfo{author}{\bibfnamefont{L.}~\bibnamefont{{Rezzolla}}},
  \bibinfo{author}{\bibfnamefont{N.}~\bibnamefont{{Stergioulas}}},
  \bibinfo{author}{\bibfnamefont{J.~A.} \bibnamefont{{Font}}},
  \bibnamefont{and} \bibinfo{author}{\bibfnamefont{E.}~\bibnamefont{{Seidel}}},
  \bibinfo{journal}{\prd} \textbf{\bibinfo{volume}{71}},
  \bibinfo{pages}{024035} (\bibinfo{year}{2005}).

\bibitem[{\citenamefont{{Brown} et~al.}(2009)\citenamefont{{Brown}, {Diener},
  {Sarbach}, {Schnetter}, and {Tiglio}}}]{brown2009}
\bibinfo{author}{\bibfnamefont{D.}~\bibnamefont{{Brown}}},
  \bibinfo{author}{\bibfnamefont{P.}~\bibnamefont{{Diener}}},
  \bibinfo{author}{\bibfnamefont{O.}~\bibnamefont{{Sarbach}}},
  \bibinfo{author}{\bibfnamefont{E.}~\bibnamefont{{Schnetter}}},
  \bibnamefont{and} \bibinfo{author}{\bibfnamefont{M.}~\bibnamefont{{Tiglio}}},
  \bibinfo{journal}{\prd} \textbf{\bibinfo{volume}{79}},
  \bibinfo{pages}{044023} (\bibinfo{year}{2009}).

\bibitem[{\citenamefont{{Shibata} and {Nakamura}}(1995)}]{shibata1995}
\bibinfo{author}{\bibfnamefont{M.}~\bibnamefont{{Shibata}}} \bibnamefont{and}
  \bibinfo{author}{\bibfnamefont{T.}~\bibnamefont{{Nakamura}}},
  \bibinfo{journal}{\prd} \textbf{\bibinfo{volume}{52}}, \bibinfo{pages}{5428}
  (\bibinfo{year}{1995}).

\bibitem[{\citenamefont{{Baumgarte} and {Shapiro}}(1999)}]{baumgarte1999}
\bibinfo{author}{\bibfnamefont{T.~W.} \bibnamefont{{Baumgarte}}}
  \bibnamefont{and} \bibinfo{author}{\bibfnamefont{S.~L.}
  \bibnamefont{{Shapiro}}}, \bibinfo{journal}{\prd}
  \textbf{\bibinfo{volume}{59}}, \bibinfo{pages}{024007}
  (\bibinfo{year}{1999}).

\bibitem[{\citenamefont{{Giacomazzo} and {Rezzolla}}(2007)}]{giacomazzo2007}
\bibinfo{author}{\bibfnamefont{B.}~\bibnamefont{{Giacomazzo}}}
  \bibnamefont{and}
  \bibinfo{author}{\bibfnamefont{L.}~\bibnamefont{{Rezzolla}}},
  \bibinfo{journal}{Classical and Quantum Gravity}
  \textbf{\bibinfo{volume}{24}}, \bibinfo{pages}{S235} (\bibinfo{year}{2007}).

\bibitem[{\citenamefont{{M{\"o}sta} et~al.}(2014)\citenamefont{{M{\"o}sta},
  {Mundim}, {Faber}, {Haas}, {Noble}, {Bode}, {L{\"o}ffler}, {Ott}, {Reisswig},
  and {Schnetter}}}]{mosta2014}
\bibinfo{author}{\bibfnamefont{P.}~\bibnamefont{{M{\"o}sta}}},
  \bibinfo{author}{\bibfnamefont{B.~C.} \bibnamefont{{Mundim}}},
  \bibinfo{author}{\bibfnamefont{J.~A.} \bibnamefont{{Faber}}},
  \bibinfo{author}{\bibfnamefont{R.}~\bibnamefont{{Haas}}},
  \bibinfo{author}{\bibfnamefont{S.~C.} \bibnamefont{{Noble}}},
  \bibinfo{author}{\bibfnamefont{T.}~\bibnamefont{{Bode}}},
  \bibinfo{author}{\bibfnamefont{F.}~\bibnamefont{{L{\"o}ffler}}},
  \bibinfo{author}{\bibfnamefont{C.~D.} \bibnamefont{{Ott}}},
  \bibinfo{author}{\bibfnamefont{C.}~\bibnamefont{{Reisswig}}},
  \bibnamefont{and}
  \bibinfo{author}{\bibfnamefont{E.}~\bibnamefont{{Schnetter}}},
  \bibinfo{journal}{Classical and Quantum Gravity}
  \textbf{\bibinfo{volume}{31}}, \bibinfo{pages}{015005}
  (\bibinfo{year}{2014}).

\bibitem[{\citenamefont{{Hawke} et~al.}(2005)\citenamefont{{Hawke},
  {L{\"o}ffler}, and {Nerozzi}}}]{hawke2005}
\bibinfo{author}{\bibfnamefont{I.}~\bibnamefont{{Hawke}}},
  \bibinfo{author}{\bibfnamefont{F.}~\bibnamefont{{L{\"o}ffler}}},
  \bibnamefont{and}
  \bibinfo{author}{\bibfnamefont{A.}~\bibnamefont{{Nerozzi}}},
  \bibinfo{journal}{\prd} \textbf{\bibinfo{volume}{71}},
  \bibinfo{pages}{104006} (\bibinfo{year}{2005}).

\bibitem[{\citenamefont{{Bona} et~al.}(1995)\citenamefont{{Bona}, {Mass{\'o}},
  {Seidel}, and {Stela}}}]{bona1995}
\bibinfo{author}{\bibfnamefont{C.}~\bibnamefont{{Bona}}},
  \bibinfo{author}{\bibfnamefont{J.}~\bibnamefont{{Mass{\'o}}}},
  \bibinfo{author}{\bibfnamefont{E.}~\bibnamefont{{Seidel}}}, \bibnamefont{and}
  \bibinfo{author}{\bibfnamefont{J.}~\bibnamefont{{Stela}}},
  \bibinfo{journal}{Physical Review Letters} \textbf{\bibinfo{volume}{75}},
  \bibinfo{pages}{600} (\bibinfo{year}{1995}).

\bibitem[{\citenamefont{{Bardeen}}(1980)}]{bardeen1980}
\bibinfo{author}{\bibfnamefont{J.~M.} \bibnamefont{{Bardeen}}},
  \bibinfo{journal}{\prd} \textbf{\bibinfo{volume}{22}}, \bibinfo{pages}{1882}
  (\bibinfo{year}{1980}).

\bibitem[{\citenamefont{{Mukhanov} et~al.}(1992)\citenamefont{{Mukhanov},
  {Feldman}, and {Brandenberger}}}]{mukhanov1992}
\bibinfo{author}{\bibfnamefont{V.~F.} \bibnamefont{{Mukhanov}}},
  \bibinfo{author}{\bibfnamefont{H.~A.} \bibnamefont{{Feldman}}},
  \bibnamefont{and} \bibinfo{author}{\bibfnamefont{R.~H.}
  \bibnamefont{{Brandenberger}}}, \bibinfo{journal}{\physrep}
  \textbf{\bibinfo{volume}{215}}, \bibinfo{pages}{203} (\bibinfo{year}{1992}).

\bibitem[{\citenamefont{{Bertschinger}}(2011)}]{bertschinger2011}
\bibinfo{author}{\bibfnamefont{E.}~\bibnamefont{{Bertschinger}}},
  \bibinfo{journal}{Philosophical Transactions of the Royal Society of London
  Series A} \textbf{\bibinfo{volume}{369}}, \bibinfo{pages}{4947}
  (\bibinfo{year}{2011}).

\bibitem[{\citenamefont{{Sachs} and {Wolfe}}(1967)}]{sachs1967}
\bibinfo{author}{\bibfnamefont{R.~K.} \bibnamefont{{Sachs}}} \bibnamefont{and}
  \bibinfo{author}{\bibfnamefont{A.~M.} \bibnamefont{{Wolfe}}},
  \bibinfo{journal}{\apj} \textbf{\bibinfo{volume}{147}}, \bibinfo{pages}{73}
  (\bibinfo{year}{1967}).

\bibitem[{\citenamefont{{Ballesteros} et~al.}(2012)\citenamefont{{Ballesteros},
  {Hollenstein}, {Jain}, and {Kunz}}}]{ballesteros2012}
\bibinfo{author}{\bibfnamefont{G.}~\bibnamefont{{Ballesteros}}},
  \bibinfo{author}{\bibfnamefont{L.}~\bibnamefont{{Hollenstein}}},
  \bibinfo{author}{\bibfnamefont{R.~K.} \bibnamefont{{Jain}}},
  \bibnamefont{and} \bibinfo{author}{\bibfnamefont{M.}~\bibnamefont{{Kunz}}},
  \bibinfo{journal}{\jcap} \textbf{\bibinfo{volume}{5}}, \bibinfo{pages}{038}
  (\bibinfo{year}{2012}).

\bibitem[{\citenamefont{{Vulcanov} and {Alcubierre}}(2002)}]{vulcanov2002}
\bibinfo{author}{\bibfnamefont{D.~N.} \bibnamefont{{Vulcanov}}}
  \bibnamefont{and}
  \bibinfo{author}{\bibfnamefont{M.}~\bibnamefont{{Alcubierre}}},
  \bibinfo{journal}{International Journal of Modern Physics C}
  \textbf{\bibinfo{volume}{13}}, \bibinfo{pages}{805} (\bibinfo{year}{2002}).

\bibitem[{\citenamefont{{Friedmann}}(1922)}]{friedmann1922}
\bibinfo{author}{\bibfnamefont{A.}~\bibnamefont{{Friedmann}}},
  \bibinfo{journal}{Zeitschrift fur Physik} \textbf{\bibinfo{volume}{10}},
  \bibinfo{pages}{377} (\bibinfo{year}{1922}).

\bibitem[{\citenamefont{{Friedmann}}(1924)}]{friedmann1924}
\bibinfo{author}{\bibfnamefont{A.}~\bibnamefont{{Friedmann}}},
  \bibinfo{journal}{Zeitschrift fur Physik} \textbf{\bibinfo{volume}{21}},
  \bibinfo{pages}{326} (\bibinfo{year}{1924}).

\bibitem[{\citenamefont{{Planck Collaboration}
  et~al.}(2015)\citenamefont{{Planck Collaboration}, {Ade}, {Aghanim},
  {Arnaud}, {Ashdown}, {Aumont}, {Baccigalupi}, {Banday}, {Barreiro},
  {Bartlett} et~al.}}]{planck2015params}
\bibinfo{author}{\bibnamefont{{Planck Collaboration}}},
  \bibinfo{author}{\bibfnamefont{P.~A.~R.} \bibnamefont{{Ade}}},
  \bibinfo{author}{\bibfnamefont{N.}~\bibnamefont{{Aghanim}}},
  \bibinfo{author}{\bibfnamefont{M.}~\bibnamefont{{Arnaud}}},
  \bibinfo{author}{\bibfnamefont{M.}~\bibnamefont{{Ashdown}}},
  \bibinfo{author}{\bibfnamefont{J.}~\bibnamefont{{Aumont}}},
  \bibinfo{author}{\bibfnamefont{C.}~\bibnamefont{{Baccigalupi}}},
  \bibinfo{author}{\bibfnamefont{A.~J.} \bibnamefont{{Banday}}},
  \bibinfo{author}{\bibfnamefont{R.~B.} \bibnamefont{{Barreiro}}},
  \bibinfo{author}{\bibfnamefont{J.~G.} \bibnamefont{{Bartlett}}},
  \bibnamefont{et~al.}, \bibinfo{journal}{ArXiv e-prints}
  (\bibinfo{year}{2015}).

\bibitem[{\citenamefont{{Bennett} et~al.}(2013)\citenamefont{{Bennett},
  {Larson}, {Weiland}, {Jarosik}, {Hinshaw}, {Odegard}, {Smith}, {Hill},
  {Gold}, {Halpern} et~al.}}]{bennett2013}
\bibinfo{author}{\bibfnamefont{C.~L.} \bibnamefont{{Bennett}}},
  \bibinfo{author}{\bibfnamefont{D.}~\bibnamefont{{Larson}}},
  \bibinfo{author}{\bibfnamefont{J.~L.} \bibnamefont{{Weiland}}},
  \bibinfo{author}{\bibfnamefont{N.}~\bibnamefont{{Jarosik}}},
  \bibinfo{author}{\bibfnamefont{G.}~\bibnamefont{{Hinshaw}}},
  \bibinfo{author}{\bibfnamefont{N.}~\bibnamefont{{Odegard}}},
  \bibinfo{author}{\bibfnamefont{K.~M.} \bibnamefont{{Smith}}},
  \bibinfo{author}{\bibfnamefont{R.~S.} \bibnamefont{{Hill}}},
  \bibinfo{author}{\bibfnamefont{B.}~\bibnamefont{{Gold}}},
  \bibinfo{author}{\bibfnamefont{M.}~\bibnamefont{{Halpern}}},
  \bibnamefont{et~al.}, \bibinfo{journal}{\apj} \textbf{\bibinfo{volume}{208}},
  \bibinfo{pages}{20} (\bibinfo{year}{2013}).

\bibitem[{\citenamefont{{Daverio} et~al.}(2016)\citenamefont{{Daverio},
  {Dirian}, and {Mitsou}}}]{daverio2016}
\bibinfo{author}{\bibfnamefont{D.}~\bibnamefont{{Daverio}}},
  \bibinfo{author}{\bibfnamefont{Y.}~\bibnamefont{{Dirian}}}, \bibnamefont{and}
  \bibinfo{author}{\bibfnamefont{E.}~\bibnamefont{{Mitsou}}},
  \bibinfo{journal}{ArXiv e-prints}  (\bibinfo{year}{2016}),
  \eprint{1611.03437}.

\end{thebibliography}

\end{document}